\documentclass[preprint]{aastex}
\usepackage{graphicx}
\usepackage{longtable}
\usepackage{amssymb}
\usepackage{color}

\makeatletter

\newcommand{\Rmnum}[1]{\expandafter\@slowromancap\romannumeral #1@}
\makeatother

\begin{document}

\title{``Invisible AGN'' \Rmnum{1}: Sample Selection and Optical/Near-IR Spectral Energy Distributions}
\author{Ting Yan, John T. Stocke, Jeremy Darling}
\affil{Center for Astrophysics and Space Astronomy, Department of Astrophysical and Planetary Sciences, UCB 389,
University of Colorado, Boulder, CO 80309-0389, USA; \email{tyan@colorado.edu}}
\and
\author{Fred Hearty}
\affil{Department of Astronomy, University of Virginia, PO Box 400325, Charlottesville, VA 22904-4325, USA}

\begin{abstract}

  In order to find more examples of the elusive high-redshift molecular absorbers, we have embarked on a systematic discovery program for highly obscured, radio-loud ``invisible AGN'' using the VLA Faint Images of the Radio Sky at Twenty centimeters (FIRST) radio survey in conjunction with Sloan Digital Sky Survey (SDSS) to identify 82 strong ($\geq$ 300 mJy) radio sources positionally coincident with late-type, presumably gas-rich galaxies. In this first paper, the basic properties of this sample are described including the selection process and the analysis of the spectral-energy-distributions (SEDs) derived from the optical (SDSS) $+$ near-IR (NIR) photometry obtained by us at the Apache Point Observatory 3.5m. The NIR images confirm the late-type galaxy morphologies found by SDSS for these sources in all but a few (6 of 70) cases (12 previously well-studied or misclassified sources were culled). Among 70 sources in the final sample, 33 show galaxy type SEDs, 17 have galaxy components to their SEDs, and 20 have quasar power-law continua. At least 9 sources with galaxy SEDs have $K$-band flux densities too faint to be giant ellipticals if placed at their photometric redshifts. Photometric redshifts for this sample are analyzed and found to be too inaccurate for an efficient radio-frequency absorption line search; spectroscopic redshifts are required. A few new spectroscopic redshifts for these sources are presented here but more will be needed to make significant progress in this field. Subsequent papers will describe the radio continuum properties of the sample and the search for redshifted \ion{H}{1} 21~cm absorption.

\end{abstract}

\keywords{galaxies: photometry --- infrared: galaxies --- quasars: absorption lines}

\section{Introduction}

While most strong radio sources are identified with luminous elliptical galaxies or optical/UV point sources (quasars), a small number of these sources lie behind obscuring screens of gas and dust becoming invisible to us at optical/UV and possibly also at near-IR wavelengths.  Identifying highly obscured, ``invisible AGN'' can lead to the detection and study of highly-redshifted radio-frequency atomic and molecular absorption line systems, which can provide important information about the early universe unobtainable in other ways. Radio absorption lines provide unparalleled velocity resolution and detailed physical diagnostics. And, where the background radio source is spatially-resolved, very-long baseline interferometry maps in the \ion{H}{1} absorption line can provide the best spatial resolution ($\sim$ 10 pc at $z$=1--2) of the gas distribution in distant galaxies; magnetic field structures also can be mapped using Faraday rotation measures of a polarized background source \citep{gae07}, or using Zeeman splitting in the 21 cm line itself in the future. Atomic plus molecular absorption can provide much additional information. First, molecules trace the gas that supplies star formation so that the strengths and ratios of their absorption lines can be used to determine the physical conditions (e.g., temperature, pressure, density, chemical abundance and magnetic field strength) of the locations where most stars have formed over cosmic time. Second, the combination of various \ion{H}{1} and molecular absorption/emission lines makes it possible to probe whether and by how much dimensionless fundamental physical constants have changed as the universe evolves.

So far there are about 80 \ion{H}{1} 21~cm absorbers detected at $z>0.1$. Only five of them are molecular absorbers\footnotemark. The first system, PKS~1413+135, was detected in \ion{H}{1} absorption in 1992 at $z = 0.25$ (Carilli, Perlman \& Stocke 1992), in CO absorption in 1994 (Wiklind \& Combes) and subsequently in nearly a dozen species (Wiklind \& Combes 1997). This source is a BL Lac object residing in an edge-on spiral galaxy and the nucleus is highly obscured (A$_{\textrm {\scriptsize V}} >$ 20; Stocke et al.1992; Carllli et al. 1992). Then B3~1504+377 was found at $z=0.67$ (Wiklind \& Combes 1996a), which is also an intrinsic molecular absorption system. The other three molecular absorbers are all in gravitational lens systems: B0218+357 at $z_{\textrm{\scriptsize em}}=0.94$ with the lensing object at $z_{\textrm{\scriptsize abs}}=0.69$ (Wiklind \& Combes 1995, Menten \& Reid 1996, Gerin et al. 1997), PKS 1830-211 at $z_{\textrm{\scriptsize em}}=2.5$ with the lensing object at $z_{\textrm{\scriptsize abs}}=0.89$ (Wiklind \& Combes 1996b) and PMN J0134-0931 at $z_{\textrm{\scriptsize em}}=2.2$ with the lensing object at $z_{\textrm{\scriptsize abs}}=0.77$ (Kanekar et al. 2005, 2012). OH lines are detected in all five systems but only two have conjugate OH molecular absorption/emission lines (PKS 1413+135, \citealt{dar04}, \citealt{kan04}; PMN J0134-0931, \citealt{kan12}) important for fundamental constants work \citep{dar03}.

\footnotetext{In this paper molecular absorption refers to the multitude of species seen in the radio-frequency spectra of giant molecular clouds, not the UV absorption transitions of H$_2$ and CO absorption arising in cool halo gas probed by far-UV spectroscopy in the Milky Way and high-$z$ QSO spectra of intervening galaxies (\citealt{not10}, \citealt{sri08})}

Although ``blind'' searches of samples of strong radio sources have been undertaken to find molecular absorption, mostly they have led to negative results, although a few new \ion{H}{1} 21~cm absorbers have been found (\citealt{cur11a}, \citealt{gup09}). Some of the unsuccessful surveys for high-$z$ molecular absorbers include: millimeter observations of strong sources with known redshifts (Willett, Kanekar, \& Carilli, private communication), cm- and m-wave observations of reddened quasars and radio galaxies, sources with damped Ly$\alpha$ absorbers, and type-2 quasars, etc. (Curran et al. 2006, 2011; \citealt{cur10}). While a few new \ion{H}{1} 21~cm absorbers have been discovered, no new molecular absorbers have been discovered in these searches. This suggests both that high-z molecular absorbers are very rare and that rather standard techniques cannot be used to discover molecular absorbers (e.g., searches using samples containing damped Ly $\alpha$ absorbers; \citealt{gup09}); i.e., the obscuration we are trying to find significantly absorbs and reddens the quasar optical-UV continuum so that it is not detectable in most cases. Thus, optically-bright quasars are not a good sample to use to search for radio-frequency molecular absorption \citep{cur08, cur11b}.

With the overall goal of discovering new examples of high-$z$ molecular absorbers, we have embarked on an extensive discovery program using a new selection method devised specifically for this task. In this paper we describe the first steps towards the ultimate goal of discovering a substantial sample of high-$z$ radio absorption line systems. This paper is arranged as follows: In Section 2 we describe this new selection method which combines Very Large Array (VLA) Faint Images of the Radio Sky at Twenty centimeters (FIRST; \citealt{bec95}) radio source detections with Sloan Digital Sky Survey (SDSS; \citealt{yor00}) photometry and morphologies for the coincident galaxies designed to select strong radio sources which are associated with late-type galaxies. In Section 3 we describe new broad-band near-IR (NIR) imaging of this sample obtained by us at the Apache Point Observatory's 3.5m telescope (APO hereafter) which allows us to search for nuclear point sources which are obscured optically and to extend the spectral energy distributions (SEDs) of these sources from the SDSS optical photometry to longer wavelengths. The analysis of the NIR images and of the optical/NIR SEDs is described in Section 4. A few new spectroscopic redshifts obtained in the progress of this work are also reported in this Section.  Individual results for some representative and also unusual objects are presented in Section 5 and candidates for finding atomic and molecular absorption are identified. We conclude with a summary of the observational results and discuss the future work using our on-going survey in Section 6.

Paper 2 in this series will describe high-resolution VLA and Very Long Baseline Array (VLBA) radio maps of these sources to identify those sources containing very compact structures and to identify sources which have radio/optical-NIR position offsets indicative of foreground, not associated absorption. A first round of \ion{H}{1} 21~cm absorption observations from this sample will be presented in Paper 3.

\section{Sample Selection}

\subsection{Background}

The host galaxies of radio-loud AGN are almost exclusively giant ellipticals (e.g. Urry \& Padovani 1995, McLeod 2006). This is true for all the subclasses of radio-louds including quasars, FRI and II type radio galaxies and BL Lac Objects. While we might generically expect radio sources associated with a spiral or irregular host galaxy or with a merger to possess foreground absorption, their occurrence is extremely rare. But recent large surveys allow us to find rare objects that do not obey the general tendency for radio-loud AGN to be hosted by giant ellipticals for specific reasons.  For example, Wilson \& Colbert (1995) have suggested that the origin of radio-loud AGN is in the merger of two supermassive Black Holes (SMBHs) when two disk systems merge to form an elliptical galaxy (Mihos \& Hernquist 1996).  If the Wilson \& Colbert (1995) scenario is correct, the merger of the SMBHs first creates a luminous but small-scale ($\sim$ 1kpc) radio-loud AGN termed a ``Compact Symmetric Object'' (CSO; \citealt{beg96}; \citealt{per01}) which can then evolve into a large-scale classical double (FR II) through subsequent outbursts. And if the CSO forms quite quickly after the merger, it can be accompanied by remnant gas and dust from the merger \citep{wil10}. This circumstance can cause the galaxy's optical morphology to appear distorted and provide gas for foreground absorption at the same time. Indeed, some CSOs have disturbed optical morphologies \citep{per01} and a high fraction ($\sim$ 20--40\%) of compact radio sources including CSOs have \ion{H}{1} absorption in their radio spectra (\citealt{van89,ver03,gup06,cha11}; but see \citealt{cur10}). PKS~1413+135, the first molecular absorber detected, is also a CSO \citep{per96} with intrinsic absorption.

Besides the two intrinsic molecular absorbers, known gravitational lenses have provided 3 other known molecular absorbers, B2~0218+357, PKS~1830-211 and PMN J0134-0934. Gravitational lenses possess unusual optical morphology due both to their multiple images and to the presence of a foreground galaxy. It is also possible that a foreground, gas-rich galaxy can produce atomic and/or molecular absorption even without producing multiple images (Narayan \& Schneider 1990). Again, a correlation between unusual optical morphology and foreground absorption can result, although in this case the absorbing gas is external to the AGN and its host.

Another rare class of radio-loud AGN which possesses unusual optical morphology is the so-called ``alignment effect'' radio galaxy (2 known examples are in our sample) in which UV emission lines from the gas surrounding the radio jets are redshifted to the optical and are responsible for a non-elliptical morphology in certain wavebands (e.g., Zirm, Dickinson \& Dey 2003). This circumstance has not led to the discovery of molecular absorbers, probably due to the very extended radio source structure required to create the ``alignment effect''. On the other hand, the first two classes described above often include radio sources with compact radio structure. Compact structure facilitates the detection of atomic and molecular absorption because the absorption detection does not get diluted by flux which does not go through the absorbing gas (see Paper 2).

\subsection{The Radio-loud Late-type Galaxy Sample}

The 1.4 GHz VLA FIRST survey covers over 9,000 square degrees of the North and South Galactic Caps. FIRST has a typical flux density rms of 0.15 mJy, and a resolution of 5$^{\prime\prime}$. The Data Release 5 (DR5) of SDSS covers about 8,000 square degrees of almost exactly the same area of the sky as FIRST. This field includes approximately 1500 FIRST sources with flux densities $>300 $ mJy. Among these sources, 808 are associated with SDSS sources, i.e., one or more optical objects are detected within 1.5$^{\prime\prime}$ of the radio centroid; thus, it is possible that some of the associated SDSS sources are {\bf not} the AGN host galaxy. Excluding 467 sources classified as stars which might actually be QSOs, 341 are probably the host galaxies of the radio sources. Note that star-galaxy separation is robust down to $r \sim 21.5$ mag in the SDSS \citep{lup01}. To select non-elliptical galaxies, two standards are used. The first is that the best-fit of the radial profile is exponential (typical for spiral galaxies) rather than De Vaucouleurs (typical for elliptical galaxies; 94 sources). The other is that the core compactness parameter\footnotemark~is smaller than 2.6 (also typical for spiral galaxies; 200 sources). These two criteria also eliminate quasars and point-source (AGN) dominated galaxies as well. The combination of the two standards makes the selection fairly reliable (Strateva et al. 2001) but there could still be some mis-classified cases. All SDSS classifications employed the $i$-band images (rest-frame $g$-band at the median photometric redshift of the sample, $z\approx$ 0.6), which typically have the highest signal-to-noise images for our sample.

\footnotetext{The ratio of the Petrosian radii containing 90\% and 50\% of the galaxy light. Same as the the concentration index defined in \citet{str01}.}

Overall, 82 objects in SDSS DR5 meet both these criteria (see Table 1). None of these sources has more than one optical counterpart, although the possibility remains that some of the optical sources are actually merged images of two or more galaxies. Because we were interested in obtaining a representative sample of these rare AGN, we did not conduct any completeness tests for this sample, nor did we obtain a new sample using later SDSS data releases. Two nearby galaxies are excluded because their intrinsic radio luminosities are much lesser than typical radio-loud AGN (J1413$-$0312 = NGC 5506 and J1352$+$3126 = UGC 8782). Also we removed J0834$+$1700 because the SDSS galaxy is coincident with one of the two double lobes in a classical double radio source; an unobscured elliptical galaxy is located between the double lobes, and so is the more likely counterpart to the radio source. We have also excluded 4 already well-studied objects: PKS 1413+135 (J1415$+$1320) mentioned above, another well-known low-$z$ CSO 4C $+$12.50 (J1347$+$1217), and the ``alignment effect'' radio galaxies, 4C +23.27 (J1120+2327, @ z=1.819) and 4C+24.28 (J1348$+$2415, @ $z$=2.879), which lack core radio emission. Our sample selection is demonstrably successful because all the three possible situations mentioned above have a few known examples in the sample. PKS 1413+135 is the first discovered high-z molecular absorber and a CSO. B2~0748+27 (J0751$+$2716) is a gravitational lens (Lehar et al. 1997).  And 4C +23.27 is the well-known ``alignment effect'' radio galaxy 3C~256 (Zirm, Dickinson \& Dey, 2003) just mentioned. There are a couple of other known CSOs and Compact Steep Spectrum Sources (CSSs; slightly larger in physical extent of 1--20 kpc but similar to CSOs; \citealt{dev98}) present and more have been found in the course of this work (Paper 2). A few host galaxies appear to be merging with another galaxy.

The full list of all 82 sources in our sample is found in Table 1. This Table includes by columns: (1) coordinate style designation for the source which is used in this paper; (2) the commonly used radio source name for the source; (3) the 20 cm continuum flux density from FIRST in Jy; (4) the associated SDSS name for the optical counterpart which also provides an accurate (RA,DEC) centroid position; (5) the SDSS $r$-band ``model magnitude'' and its associated error \citep{sto02}, which results from an extrapolation of the best-fit (for this sample) exponential radial profile; (6) the photometric redshift and error from Data Release 8 (DR8) using an empirical method called ``kd-tree nearest neighbor fit'' (Csabai et al. 2007); and (7) the spectroscopic redshift where available. The spectroscopic redshifts come from references in the NASA/NED database except for the cases noted with an asterisk; these new spectroscopic redshifts were obtained by us and are described in Section 3.

Notice that most of these sources are quite faint in the optical ($\sim$ 75\% fainter than r=21) so that the application of the Strateva et al. (2001) morphology discriminators is not without some uncertainty. For this and other reasons we sought to confirm the diffuse nature of the potential invisible AGN hosts using more recent SDSS results (DR8) and new near-IR imaging (see Section 3). Since the original galaxy classification was made using SDSS DR5, we have revisited the sample selection process using DR8. In almost all cases the original classification is confirmed but in a few cases (5 in number) the DR8 compactness parameter ($C$) exceeds the $C<$ 2.60 cut suggested by Strateva et al. (2001) for late-type systems. In one other case (J1421$-$0246) the r-band SDSS galaxy image is much better fit by a De Vaucouleurs r$^{1/4}$-law than an exponential disk. Our near-IR imaging of these sources confirms these reclassifications, except in one case. For J0824+5413, our near-IR imaging reveals a very diffuse galaxy, at odds with the C=2.9 compactness value for this object in DR8. Therefore, we have retained this source in our sample. The other reclassifications are marked with daggers ($\dagger$) in Table 1 and removed from the final sample listed in Table 2.

\section{Observations}

\subsection{Near-IR Imaging}

In order to check the SDSS galaxy classifications, to complement the SDSS photometry (and thus to provide an optical/NIR SED), and to search for an extincted point-source indicative of an imbedded AGN, we obtained NIR photometry of this sample using the Near-Infrared Camera \& Fabry-Perot Spectrometer (NIC-FPS; Hearty et al. 2005) in its imaging mode on the APO 3.5 m.  In the last several years almost all of the objects in this sample were observed in the $J$ and $K_{{\textrm {\scriptsize short}}}$ bands with all objects detected in $K_{{\textrm {\scriptsize short}}}$ and most in $J$-band. $H$-band observations were made for fewer of these sources, particularly where the $H$-band photometry was important in evaluating the nature of the source. During these observations, the seeing varied from sub-1$^{\prime\prime}$ to 2$^{\prime\prime}$ with a median seeing value of $\sim 1.3^{\prime\prime}$. This image quality is approximately the same or, in many cases, better than the image quality for the SDSS.

The images were reduced in the standard manner with flat-field and dark images obtained the same night. Because all of these images were obtained using a 5-point dither pattern on the sky, the science images were used to create a sky-flat by median-filtering. Following flat-fielding the images were aligned and stacked; astrometric positions for source centroids were set to the World Coordinate System using 5-10 SDSS stars identified in these images.  Most of the objects in this project are faint, so, in order to reduce the photometric error and for consistency with the SDSS optical photometry, we first measure the magnitude in an aperture of 2$^{\prime\prime}$ ($m_{2^{\prime\prime}}$). This magnitude is similar to the fiber magnitude $m_{{\textrm {\scriptsize fiber}}}$ in SDSS (Stoughton et al. 2002). SDSS also gives the ``model magnitude'' $m_{{\textrm {\scriptsize model}}}$ as the best estimated magnitude for extended sources. Assuming an unchanged radial profile between the optical and NIR images, the ``model NIR magnitude'' is $m=m_{2^{\prime\prime}}+(m_{{\textrm {\scriptsize model}}}-m_{{\textrm {\scriptsize fiber}}})$ and is the NIR equivalent of the SDSS model magnitudes to facilitate a consistent source SED.

These are the magnitudes that we quote in Table 2 and throughout this paper. Table 2 lists our NIC-FPS photometry for the 70 sources in the trimmed sample including by column: (1) source name as in Table 1; (2--4) the next three columns list the $J$, $H$ and $K_{{\textrm {\scriptsize short}}}$ band magnitudes. The listed statistical errors include a contribution from photon statistics as well as the zero-point errors from the 5-10 2MASS stars in the field. Column (5) lists the UT date of the observations in the format YY-MM-DD; all observations of any one source were obtained sequentially on the UT date listed. Column (6) lists the classification of the source based on its optical/NIR SED. These classifications are discussed in the next Section in detail. Abbreviations include: $G$ = galaxy; $Q$ = quasar and $Q+abs$ = possible reddened quasar; $G+Q$ = galaxy $+$ quasar. Column (7) lists $z_{{\textrm {\scriptsize fit}}}$, the best-fit photometric redshifts based on optical + NIR SEDs for $G$-type sources only. Column (8) lists the Hubble type for the best-fitting SED galaxy template (see next Section for discussion) used for the classification in Column (7). The entries in the last three columns will be discussed in detail in the next Section.
 
Usually there are 5-10 non-saturated objects from the Two Micron All-Sky Survey (2MASS) catalog in the $\sim 5^{\prime}\times 5^{\prime}$ NIC-FPS field, which are used to provide zero-points for the magnitudes listed in Table 2. The magnitude error budget includes amounts from the image photon-statistics (usually $\sigma_{2^{\prime\prime}}<0.05$), the 2MASS magnitude system error ($\sigma<0.1$) and slightly larger and less well-defined errors accrued in placing the NIR magnitudes onto the SDSS ``model magnitude'' scale ($\sigma_{{\textrm {\scriptsize model}}}\sim0.1$ and $\sigma_{{\textrm {\scriptsize fiber}}}$ varies from 0.05 to 0.3). Sources which show NIR point sources, and so have more centrally-concentrated surface brightness profiles than their optical images, may have NIR ``model magnitudes'' which have been under-estimated systematically by 0.1--0.2 mags. For non-detected objects, we set a $3\sigma$ flux level recorded as a model magnitude limit in Table 2.  Blank spaces in Table 2 mean that the source was not observed in that band.

The systematic astrometric error of an SDSS frame is on the order of 50 mas per coordinate. For objects brighter than $r$ = 20 mag, the object controiding for a point source gives a random error of about 20 mas. About 5-10 SDSS stars with $r <$ 20 mag were selected in each NIR field for astrometric calibration, yielding a total positional uncertainty in the range of 50 - 100 mas. For bright FIRST radio sources the astrometric error is smaller than 50 mas so that the optical/NIR positional errors dominate the accuracy with which the SDSS, NIR and VLA/VLBA images can be registered. All sources have coincident NIR and optical positions although discrepancy exists between optical/NIR and radio positions for some of the sources mostly due to extended structure of the radio sources. More details will be discussed in Paper 2.

Specific unusual circumstances involving morphology, relative positional offsets and unusual or inconclusive 
SEDs are discussed source by source in Section 5.

Due to the faint and diffuse nature of most of the target galaxies in this sample, spectroscopic redshifts are available for only 17 of them. Originally it was thought that a reliance on the SDSS photometric redshifts would be sufficient for this program given the wide bandwidths available on current generation sub-GHz radio receivers. However, this has not proven to be the case. First, the special defining qualities of this sample make photometric redshifts (photo-$z$ hereafter) less reliable than for most galaxy samples (i.e., optical objects can be combinations of starlight plus extincted AGN). And second, the difficult radio frequency interference (RFI) environments now present at all terrestrial radio telescope sites make even typical photo-$z$ accuracy insufficient for a sensitive \ion{H}{1} and molecular line search; i.e., without an accurate spectroscopic redshift, the presence of intense, receiver-saturating RFI means that it is not possible to determine definitively whether an absorption is present without a specific frequency prediction. Due to this circumstance we initiated a program for obtaining optical and/or NIR redshifts using the spectrographs at the APO 3.5m.

\subsection{Optical and Near-IR Spectroscopy}

\begin{figure}[p]
\centering
\includegraphics{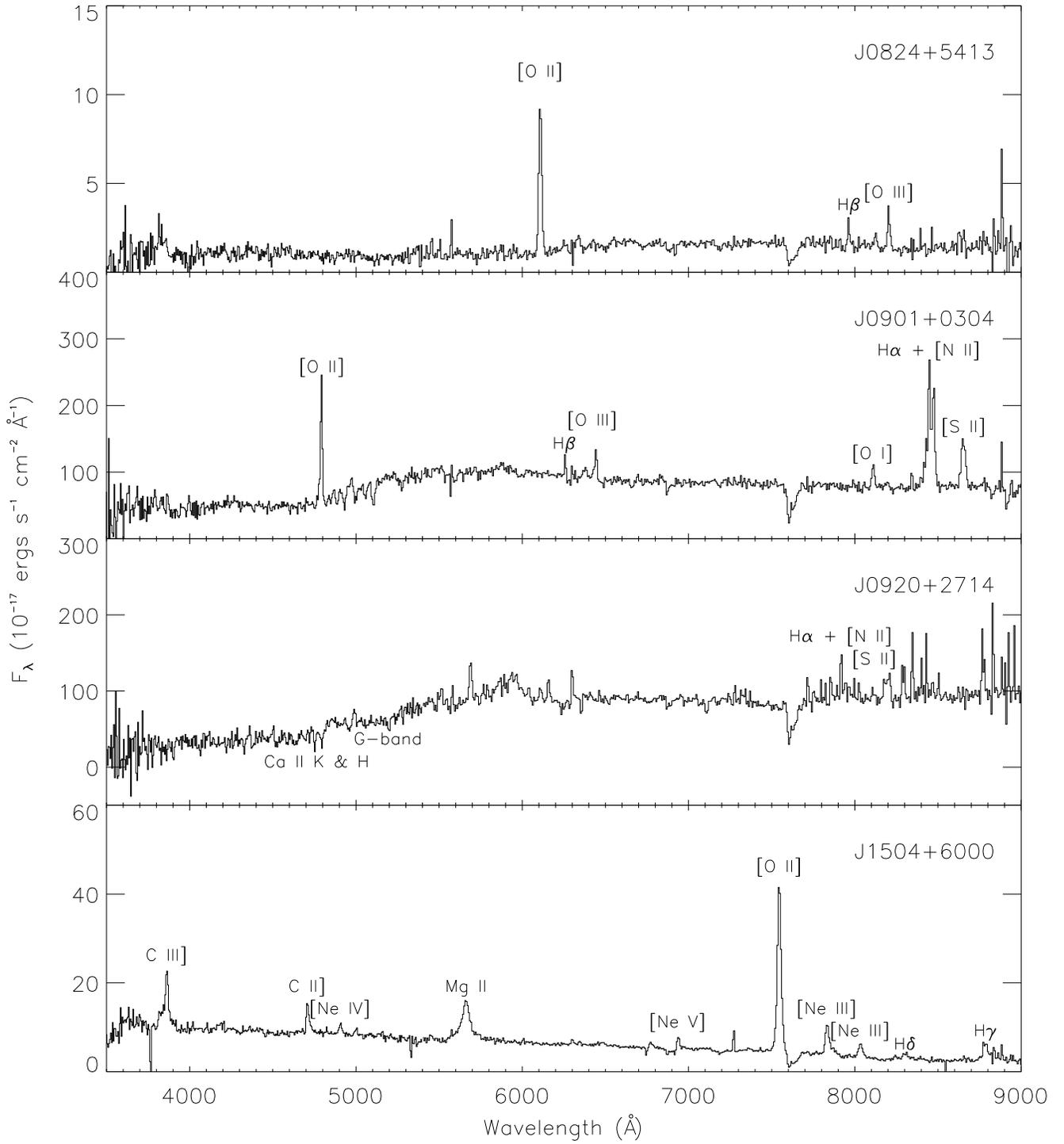}
\caption{The APO/DIS optical spectra with prominent emission/absorption lines marked.\label{fg:DIS}}\end{figure}

\begin{figure}[p]
\centering
\includegraphics{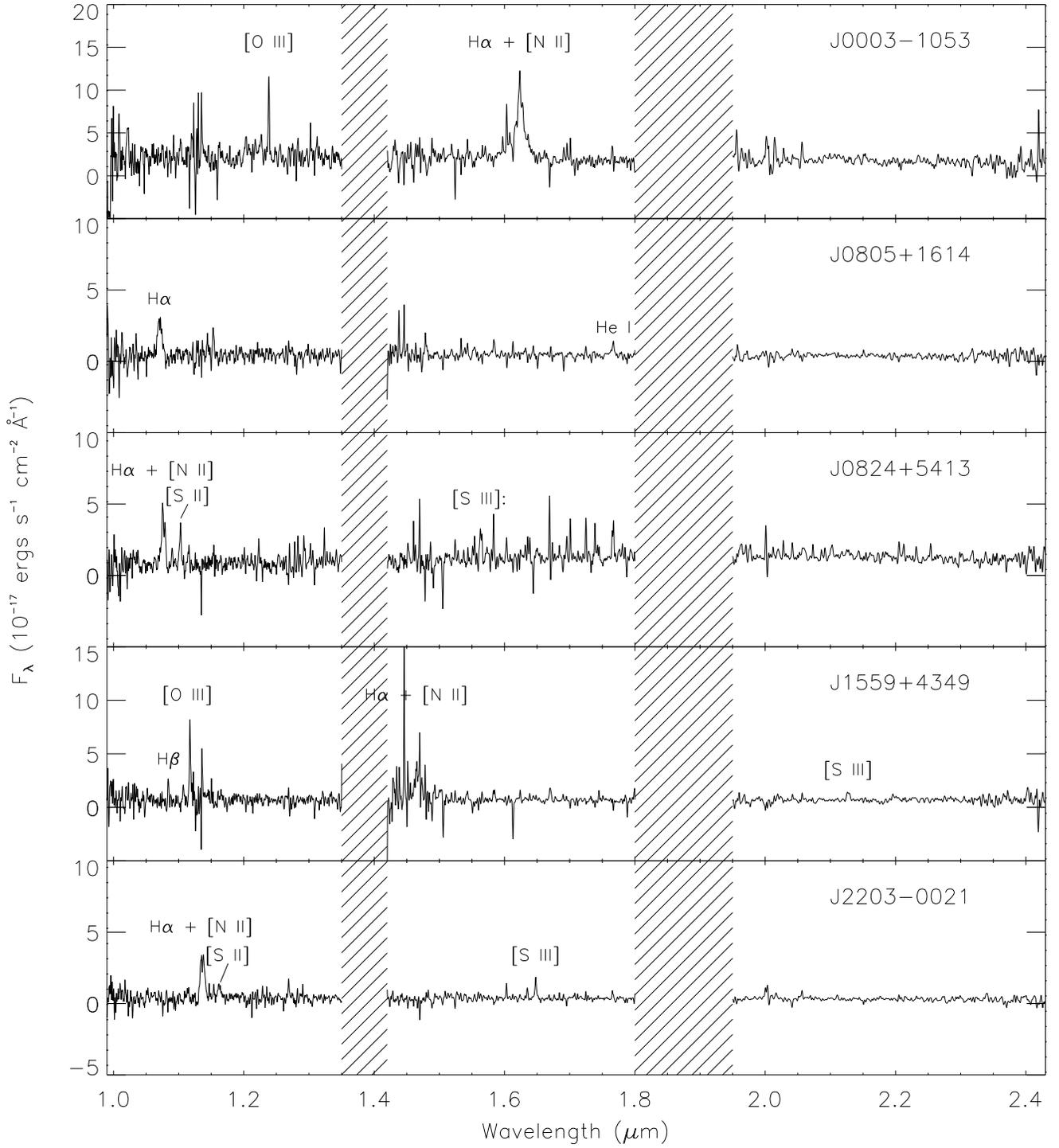}
\caption{The APO/TripleSPEC near-IR spectra with prominent emission lines marked. The shaded regions are regions of significant atmospheric absorption.\label{fg:TSPEC}}\end{figure}

The Dual Imaging Spectrograph (DIS II) was used in its standard low dispersion mode to obtain $\sim$ 150 km s$^{-1}$ resolution spectra in the 4000--9000 \AA\ wavelength region (1.8 \AA\ pix$^{-1}$ shortward of 5500~\AA\ and 2.4~\AA\ pix$^{-1}$ longward; see \citealt{oke82} for original design). TripleSpec is a cross-dispersed NIR spectrograph that provides simultaneous R $\sim$ 3500 spectra in a continuous wavelength coverage from 0.95--2.46 $\mu m$ in five spectral orders.  The instrument is described in more detail in \citet{wil04}. Where we have successfully obtained DIS II spectra, emission lines are detected and redshift errors are $\pm$30--60 km s$^{-1}$ and $\pm$90 km s$^{-1}$ where absorption and only weak emission is detected. TripleSpec NIR spectra have similar resolution but the signal-to-noise ratio is much lower so that the DIS II redshifts are preferred where we have obtained both optical and NIR spectroscopy. These new redshifts led to the detection of redshifted \ion{H}{1} 21~cm in two cases. Despite observing many more objects which possessed only photometric redshifts only one new redshifted \ion{H}{1} 21~cm absorption detection was made based on a photo-$z$ alone: J1357$+$0046 (PKS~1355+01) possesses two \ion{H}{1} components at $z_{\rm HI}$=(0.7971, 0.7962); i.e., at a significantly higher $z$ than the SDSS $z_{\textrm{{\scriptsize phot}}}$=0.57. The eight new redshifts we have obtained for this sample are at the limit of what is possible using a 4m class telescope; i.e., $r\leq20$ and/or $K\leq16.5$. The spectra we have obtained thus far for this project using DIS II and TripleSpec are listed in Table 3 with the instrument used, the redshift obtained including measurement errors and the emission lines detected. These spectra are shown in Figure \ref{fg:DIS} and Figure \ref{fg:TSPEC}.

\section{NIR Morphology, Magnitudes and Optical/NIR SEDs}

With multi-band observations, we are now able to better understand these sources despite their faintness. In this Section we use the NIR magnitudes to examine the NIR morphology of the source, use the NIR magnitudes to create an optical-NIR SED and compare these with template galaxy spectra. The $K_{{\textrm {\scriptsize short}}}$ magnitudes are then used to place these objects on the $K-z$ diagram for which many radio-loud sources show a strong correlation.

\subsection{Near-IR (NIR) Morphology}\label{sec:mor}

\begin{figure}[p]
\centering
\includegraphics{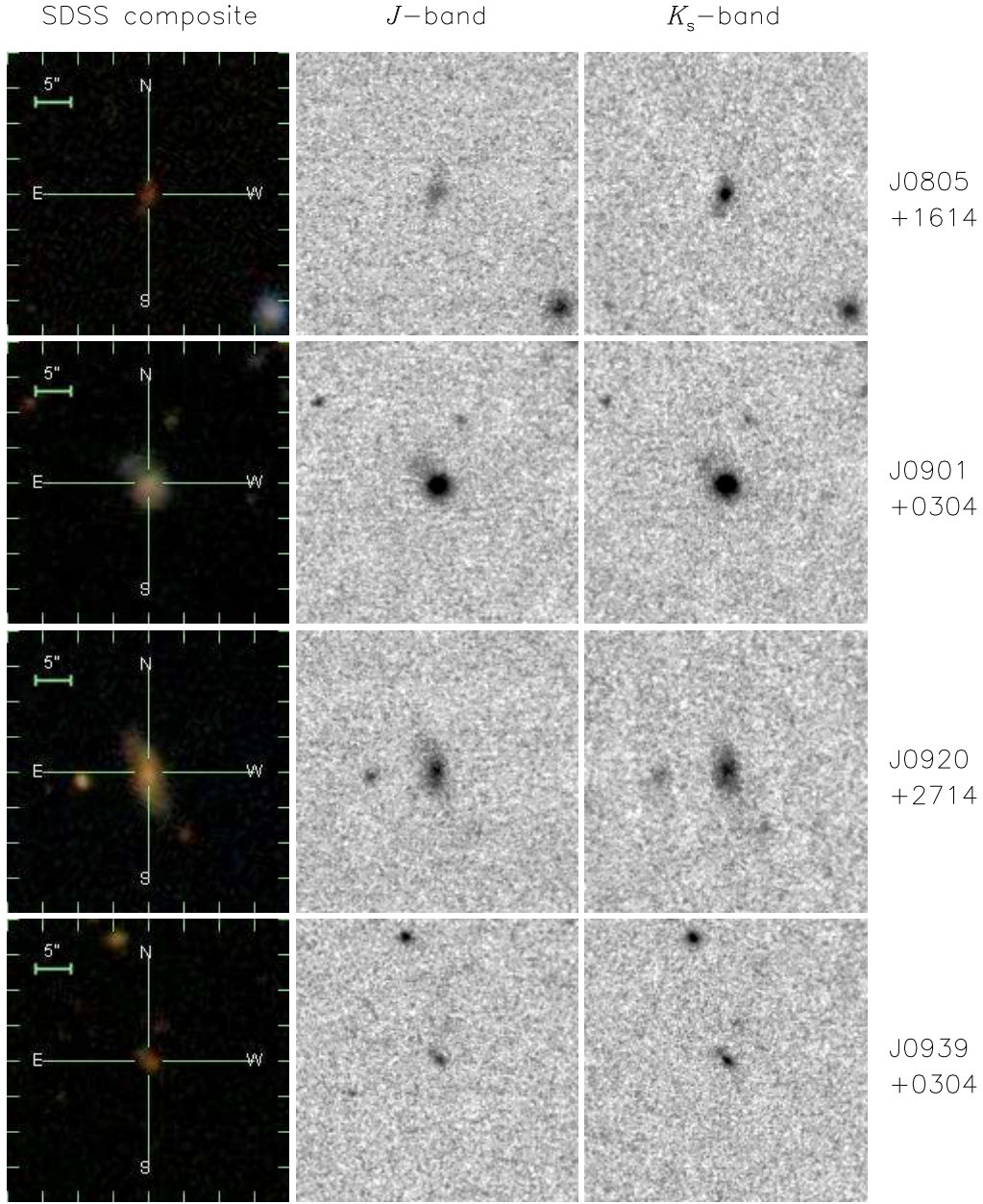}
\caption{Representative optical and NIR images from our sample all shown at the same scale and orientation as indicated
in left-hand column. Shown are the SDSS composite images ({\it left}; blue: $g$-band, green: $r$-band, red: $i$-band), and the APO 3.5m 
$J$-band ({\it middle}) and $K_{\rm s}$ band ({\it right}) images of (from top to bottom) J0805$+$1614, J0901$+$0304, J0920$+$2714, and J0939$+$0304. 
\label{fg:NIRmorph}}
\end{figure}

The NIC-FPS ($JHK$) imaging provides an independent check of the SDSS galaxy morphology classifier.  In general, these images provide ample confirmation of the galaxy morphology in that virtually all of them are diffuse sources in moderate seeing conditions (median $\sim 1.3^{\prime\prime}$, Figure \ref{fg:NIRmorph}). For a few sources, particularly some of the optically red sources, the NIR images suggest that an elliptical galaxy morphology is possible so that the Strateva et al. (2001) classification may be incorrect.  The few sources which show a centrally-concentrated structure in our NIR images are J1354$+$5650, J0749$+$2129, J0920$+$2714, J1043$+$0537, J1019$+$4408,and J0003$-$1053. These are not removed from the sample because they may still be extincted, although this is less likely compared to the more diffuse galaxies (e.g., J0920$+$2714 is detected in redshifted \ion{H}{1} 21~cm absorption). Otherwise, the optical morphology classifications for these sources made by the SDSS is confirmed.
  
\subsection{Optical-NIR Spectral Energy Distributions (SEDs)}\label{sec:SED}

\begin{figure}[p] 
\centering 
\includegraphics{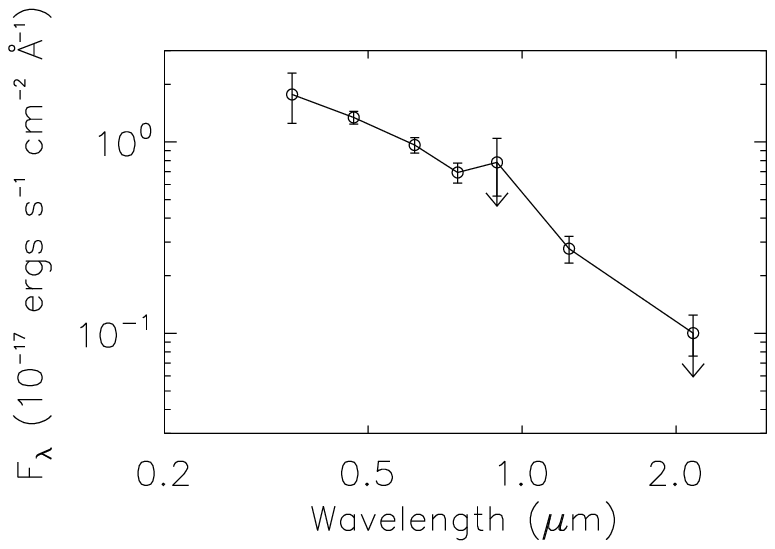} 
\includegraphics{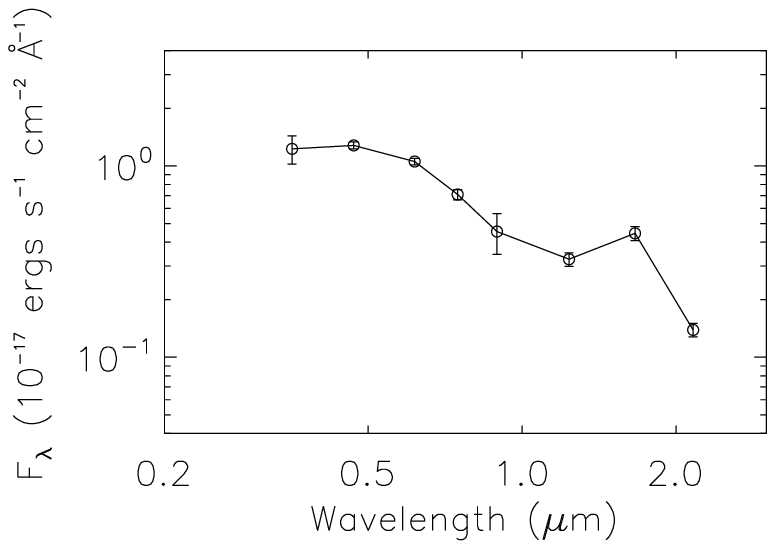} 
\caption{SED classified as $Q$ ({\it top}, J1125+1953) and $Q+abs$ ({\it bottom}, J1724+3852). \label{fg:Q}}
\end{figure}

\begin{figure}[p]
\centering
\includegraphics{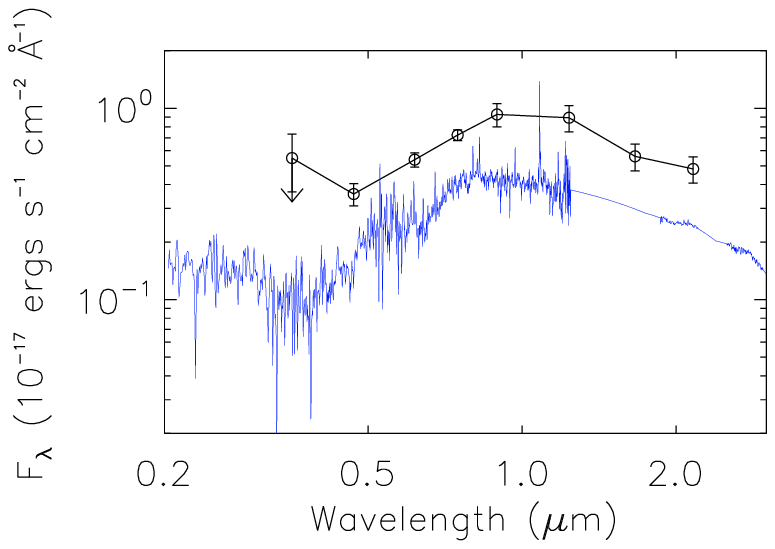}
\includegraphics{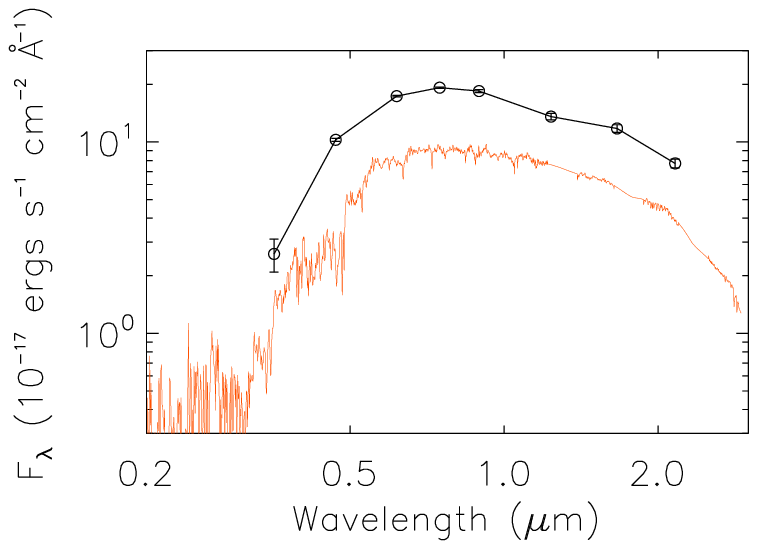}
\caption{SED classified as $G$ ({\it top}, J0951+1154; {\it bottom}, J0920+2714). The spectra are template galaxies of Sc type at $z=0.65$ ({\it top}) and S0 type at $z=0.23$ ({\it bottom}). \label{fg:G}}
\end{figure}

\begin{figure}[p]
\centering
\includegraphics{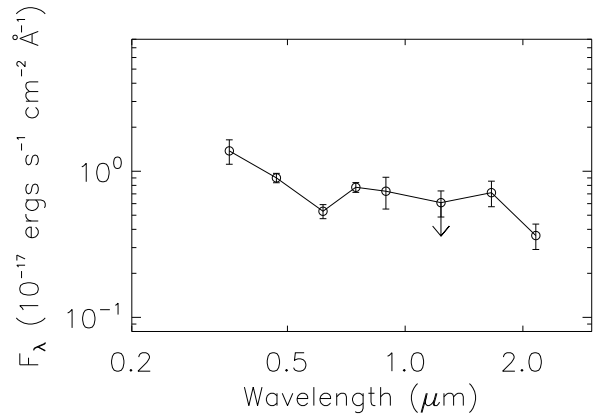}
\caption{SED classified as $G+Q$ (J0917+4725).
\label{fg:G+Q}}
\end{figure}

These sources have a wide variety of optical-NIR SEDs due to three factors: variations in host galaxy type, the relative importance of the host galaxy and the AGN as a function of wavelength, and the amount of nuclear obscuration. In an attempt to categorize their SEDs we have placed each source into one of the following three categories, listed these in Table 2 and show a few examples of each type in Figures \ref{fg:Q}--\ref{fg:G+Q}):

\begin{itemize}

\item Quasar ({\it Q} in column 6 of Table 2) or obscured quasar ({\it Q$+$abs}). The {\it Q} SED is approximated by a single power-law declining from blue to red in F$_{\lambda}$, although in some cases contributions from emission lines to specific flux bands are evident. Almost all previously known quasars in our sample have {\it Q}-type SEDs. Sources classified as {\it Q$+$abs} show evidence for a flattening in flux at the blue end. In some cases the SDSS {\it u}-band flux is only an upper limit making the classification uncertain between {\it Q$+$abs} and a galaxy {\it G}-type SED; i.e., it is uncertain whether the near-UV flux continues to decline or not. See Figure \ref{fg:Q} for examples.
 
\item Galaxy ({\it G}). This galaxy-like SED peaks in the middle bands and declines toward both the blue and the red ends. See Figure \ref{fg:G} for examples. If the turn-down at the blue-end is not explicitly observed ({\it u} and sometimes {\it g}-band fluxes are upper limits), the source SED could either be {\it Q$+$abs} or {\it G}. Sources are classified as {\it Q$+$abs} if the NIR slope is too steep to be a galaxy, otherwise these sources are classified as $G$.

\item Galaxy $+$ Quasar ({\it G$+$Q}). The SED is a power-law {\it Q} component with a bump in the middle bands which could be a galaxy {\it G} component. Sources missing the flattening at the red end are labeled as {\it G$+$Q} since the absence of a flattening or turn-down in the NIR could be due to contributions from a nuclear source that is obscured at shorter wavelengths. See Figure \ref{fg:G+Q} for an example.

\end{itemize}

Our classifications are largely confirmed by the 17 sources for which optical and/or near-IR spectroscopy is in-hand. Overall there are 33 sources in the $G$ class (including 12 SEDs best-fit by early-type galaxies and 21 by late-type galaxies), 17 in $G+Q$, 6 in $Q$ and 14 in $Q+{\it abs}$. While our NIR photometry was obtained several years after the SDSS optical photometry, we do not see strong effects in the SEDs arising from any AGN variability that may have occurred over that time period. The SDSS and NIR images (as in Figure 3) and the SEDs (as in Figure 4--6) will be shown for the entire sample in Yan (2012, PhD thesis, in preparation). Due to the strong possibility that the 33 sources in the $G$ class are actually rather simple in ther overall SEDs, in this next Section we concentrate on $G$-class sources in a further scrutiny of their photometric properties.

\subsection{Photometric and Spectroscopic Redshifts}

\begin{figure}[p]
\centering
\includegraphics{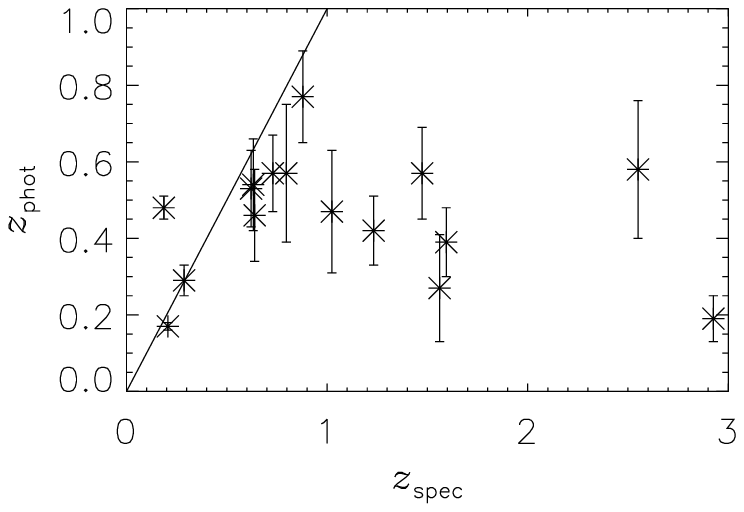}
\includegraphics{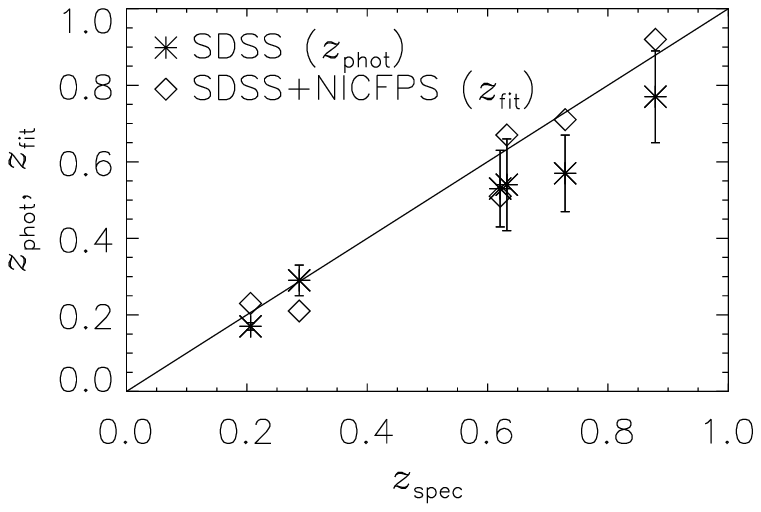} 
\caption{{\it Top} (a): Spectroscopic redshifts compared with photometric redshifts from DR8 for those 16 sources
with spectroscopic redshifts. The line shows $z_{\textrm{{\scriptsize spec}}} = z_{\textrm{{\scriptsize phot}}}$, illustrating that the fitting algorithm does not work for high-$z$ ($z \gtrsim 1$) quasars. Even for those sources not classified as quasars, the $z_{\textrm{{\scriptsize phot}}}$ is systematically lower than $z_{\textrm{{\scriptsize spec}}}$ by $\Delta z =$ 0.05--0.1 at $z\sim$ 0.5--1. {\it Bottom} (b): Photometric redshifts from DR8 (asterisks with error bars) and from our fitting routine (diamonds) using SDSS$+$NIR photometry for sources with spectroscopic redshifts and classified as $G$. See Section 4.3 for discussion.\label{fg:zspeczphot}}
\end{figure}

Since up to now this program has relied primarily on the SDSS photo-$z$s, a brief review of their basic properties is in order. Before Data Release 7 (DR7), SDSS photo-$z$s are derived from fits using typical SDSS resolution spectra of various galaxy types \citep{bud00,csa03}. A new algorithm is used in DR7 which utilizes SDSS spectroscopic catalog as a training set to determine photo-$z$s from spectroscopic objects with similar SEDs \citep{csa07}. However, objects in this training set are brighter and closer on average than typical objects in our sample. A fainter and deeper training set more representative of our sources is included in DR8 which has $r$-band magnitudes mostly between $r$ = 20--22, redshifts mostly between 0.2--0.8, and the improved average error is 0.10--0.12 (Csabai, private communication). Although these redshifts are a good potential match to our own sample, the correlation between the spectroscopic and photo-$z$s we have in-hand is less compelling. Figure \ref{fg:zspeczphot}a shows sources with known spectroscopic redshifts compared with their photometric redshifts from DR8. Only half (8 out of 16) of these sources appear reasonably well-fit by the DR8 algorithm to the 1$\sigma$ accuracy noted above. But even for the 8 with $z_{\textrm{{\scriptsize phot}}} \simeq z_{\textrm{{\scriptsize spec}}}$, the photometric redshift systematically underestimates the redshift by $\Delta z \sim$ 0.05--0.1 (see Figure \ref{fg:zspeczphot}a), i.e., the offset of \ion{H}{1} absorption can be as large as 70 MHz. We note that the only source (J1414+4554) to the left of the solid line in Figure \ref{fg:zspeczphot}a has an uncertain redshift of $z=0.186$ (Falco et al. 1998). Given its unique and unusual location in this plot we doubt the accuracy of this spectroscopic redshift (Falco et al. did not publish their spectrum) and do not use it further.

In the poorly-fit 7 sources at $z > 1$, three are classified as $Q$ SED types and 4 are classified as $Q+{\it abs}$. It is understandable in these cases that the SDSS photo-$z$ classifier obtains an unreliable redshift by trying to fit a galaxy template to a reddened quasar SED. An emission line spectrum with weak continuum may also pose a problem for the photo-$z$ algorithms. Thus, those source with $z_{\textrm{{\scriptsize spec}}} >> z_{\textrm{{\scriptsize phot}}}$ are not representative of the remainder of the sample and photo-$z$s for $Q$ and $Q+abs$  type sources in particular should not be used. 
 
\begin{figure}[p]
\centering
\includegraphics{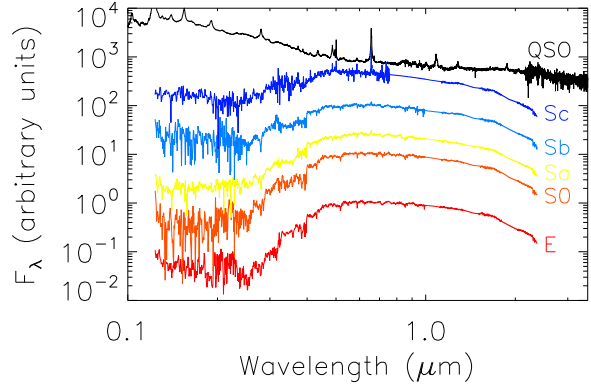}
\caption{Template spectra for typical galaxies (Mannucci et al. 2001) and unreddeded QSOs (Vanden et al. 2001, Glikman et al. 2006). \label{fg:template}}
\end{figure}

Given the potential complexities of the SEDs for these galaxies, it would be very helpful if we could use our NIR photometry to further constrain source redshifts.  But given the results shown in Figure \ref{fg:zspeczphot}a we restrict this analysis to only those sources classified as {\it G}-types in Figure \ref{fg:zspeczphot}b. Using the SED templates in Figure \ref{fg:template} we have used a least-squares approach to determine the best-fit Hubble types and redshifts ($z_{\textrm{{\scriptsize fit}}}$) for the {\it G} sources based on their SDSS $+$ NIR SEDs and listed them in Table 2 column 7 \& 8. We searched parameter space in redshift, peak flux density and discrete Hubble types (Figure \ref{fg:template}) to find the best-fit values that give the least difference between the observed flux points and the template SEDs. As shown in Figure \ref{fg:zspeczphot}b using data on {\it G}-type sources only, a photo-$z$ can be estimated to $\pm$ 0.2 at high confidence ($\sim$ 90\%). While this is true both for those sources with well-fit SDSS $+$ NIR SEDs, and for those sources with SDSS photo-{z}s only, there is a slight systematic difference between $z_{\textrm{{\scriptsize fit}}}$ and $z_{\textrm{{\scriptsize phot}}}$. This slight systematic underestimate of $\Delta z \sim$ 0.05--0.1 is the same as in Figure \ref{fg:zspeczphot}a, suggesting that the SDSS+NIR SEDs are a slightly better indication of the true redshift of these sources. However, we caution that this conclusion is based on a training set of $<$ 10 sources. Even with this improvement in the precision, photo-$z$s are inadequate for a successful redshifted H~I 21cm absorption since the above quoted precision ($\Delta z \sim 0.1$) can provide a search frequency accurate to no better than $\sim$ 40 MHz at $z\sim$ 1. Given current levels of RFI at large radio telescopes world-wide this is still inadequate for a conclusive H~I search (i.e., RFI-free regions are typically much smaller than this at $\nu <$ 1 GHz) so that spectroscopic redshifts are still required.
 
But before we can accept that the photo-$z$s for all $G$-type sources are accurate to $\pm$0.1 in $\Delta z$, we need to scrutinize them further using previous results for luminous radio galaxies. The $G$-type sources can be plotted on the well-known $K-z$ diagram that relates the $K$-magnitude and the redshift (van Breugel et al., 1998). Willott et~al. (2003) observed 49 radio galaxies with 151-MHz flux densities greater than 0.5~Jy and gave a $K-z$ relation fitted by a second-order polynomial between $K$-magnitude and $\textrm{log}_{10}z$, \begin{equation} K = 17.37+4.53\textrm{log}_{10}z-0.31(\textrm{log}_{10}z)^2.  \label{eq:K-z} \end{equation}

\begin{figure}[p]
\centering
\includegraphics{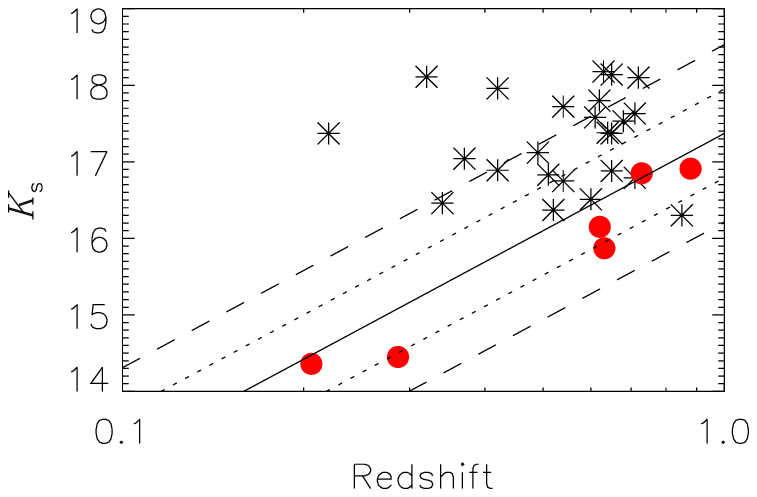}
\caption{The observed $K_{\rm s}-z$ diagram for objects whose SED is classified as $G$. The red dots represent objects with known spectroscopic redshifts. For the remainder, photometric 
redshifts are used (black asterisks). The black line is the best fit $K-z$ relation from Willott et
  al. (2003) with the dotted lines indicating the 1$\sigma$ dispersion and the dashed lines indication the 2$\sigma$ dispersion. Table 4 lists 9 sources which are fainter in $K$ than the 2$\sigma$ dispersion in the ($K-z$) relation. See Section 4.3 for discussion. \label{fg:K-z}}
\end{figure}

The dispersion in the correlation between these two quantities for radio-loud AGN is reported to be 0.58 mag up to $z=3$. If a galaxy is formed at $z>5$, undergoes an instantaneous starburst, then evolves passively and has a luminosity of $3L^*$ (typical giant elliptical galaxy hosting a radio-loud AGN) at present, the $K$-magnitude evolution of such a galaxy is close to the observed $K-z$ relation presented here. The observed $K-z$ values from our sample (excluding all $Q$ and $Q+abs$ type sources) are plotted in Figure \ref{fg:K-z} together with the $K-z$ relation from Equation \ref{eq:K-z}.

While some sample sources conform quite well to the relationship (including all those with spectroscopic redshifts), there are a significant number which should have $z\gtrsim 1$ based on their $K-$band magnitudes and instead have $z_{\textrm{{\scriptsize phot}}} <$ 1.  While we will not understand these discrepancies fully until spectroscopic redshifts are obtained, there are several possibilities including: (1) the host galaxies of some radio-loud AGN are not giant ellipticals; (2) the host galaxy, or at least the inner bulge of the host, is heavily extincted (A$_{\textrm{\scriptsize V}} > 20$) even at $K$-band (unlikely); (3) the SDSS galaxy is not the AGN host but is rather a lensing, or simply a foreground, galaxy close to the sightline; and (4) the photo-$z$s are systematically too low due to the SDSS source being a composite of quasar and galaxy. Regardless of the solution for each case, the source offset from the $K-z$ relation can be used as a diagnostic that there is some unusual circumstance for that particular source and that the photo-$z$ is unreliable. This appears to be the case for a large fraction of the sample.

Given Equation (1) and that the photo-$z$s are $<$ 1, almost all $G$-type sources in Table 2 with $K\geq$17 would be discrepant in Figure \ref{fg:K-z}. Approximately 2/3 of the sources falls into this magnitude range and so have $z_K >> z_{\textrm{{\scriptsize fit}}}$, where $z_K$ is the redshift calculated in Equation (1) from the $K$-band mags in Table 2. The 21 discrepant objects in our sample have diffuse morphology in the NIR and so are either extincted even at $K$-band, are underluminous hosts, or are foreground galaxies unrelated to the AGN. {\it Spitzer Space Observatory} mid-IR photometry of these objects can determine if they are still extincted at $K$-band. In any case these sources are good candidates for foreground absorption. In Table 4 we list those 9 sources that are more than 2$\sigma$ fainter than expected in $K$-band based on their photo-$z$s. Table 4 contains the following information by column: (1) source name; (2) $z_{\textrm{{\scriptsize fit}}}$ from Table 2 column (7); (3) Hubble type from Table 2 column (8); (4) $K$-band magnitude from Table 2 column (4); (5) $z_{\textrm{{\scriptsize K}}}$; and (6) the amount of redshift discrepancy, $\Delta z$= $z_K - z_{\textrm{{\scriptsize fit}}}$.

\section{Discussion of Selected Unusual Sources}

\hspace{5mm} \textbf{J0003$-$1053} This is one of the reddest sources in our
sample with ($r-Ks)=5.5\pm0.3$. Although the flattening of the SED longward of 1 micron is
comparable to what is expected for a $z\sim$ 2 elliptical, the observed blue flux density is much too faint
for this interpretation. So this source must be heavily obscured and is an excellent
candidate for absorption studies at $z = 1.474$.

\hspace{5mm} \textbf{J0134$+$0003}. The optical spectroscopic redshift of $z$=0.879 is from Drinkwater et al. (1997). \citet{cur06} observed this source for \ion{H}{1} absorption but severe RFI was present around the expected frequency for \ion{H}{1} at $z$=0.879.

\hspace{5mm} \textbf{J0751$+$2716} This source is a well-known ``Einstein Ring'' gravitational lens system with
$z_{{\textrm {\scriptsize lens}}}$=0.35 and $z_{{\textrm {\scriptsize QSO}}}$=3.2. While this source has been searched for redshifted \ion{H}{1} 21 cm absorption by us, RFI was present which rendered the search inconclusive.

\hspace{5mm} \textbf{J0901$+$0304} Our DIS II spectrum yields $z$ = 0.2872 based on strong narrow emission lines (Table 3; Figure \ref{fg:DIS}) and led to the discovery of 
two components of redshifted \ion{H}{1} 21~cm absorption at $z$=(0.2886, 0.2869). See Paper 3.

\hspace{5mm} \textbf{J0920$+$2714} While this source is apparently associated with a rather normal bright ($K$=14.3) galaxy at $z_{\textrm{{\scriptsize spec}}}$ = 0.2064 (see Table 3), the FIRST radio source position is offset from the SDSS galaxy position by 1.7 arcsecs, an offset we have confirmed using new VLA and VLBA maps (see Paper 2).  An \ion{H}{1} 21~cm absorption has been discovered at $z$= 0.2067 (see Paper 3), consistent with the optical/NIR galaxy; i.e., the radio-loud AGN is background to the galaxy which creates the absorption. 
 
\hspace{5mm} \textbf{J0939$+$0304} There is a point source seen only in the
$K_{\rm s}$ image of this source but not at the center of the ``host'' galaxy (Figure \ref{fg:NIRmorph}). 
Regardless of whether the AGN is in an interacting system or whether the nuclear offset is due to a
foreground galaxy, this source is an excellent candidate for absorption studies.

\hspace{5mm} \textbf{J1357$+$0046} The optical and NIR images show a faint linear string of knots, one of which is 
coincident with the radio source (see Paper 2). Despite the absence of a spectroscopic redshift, we have detected \ion{H}{1}
21~cm redshifted absorption at $z$=(0.7971, 0.7962). See Paper 3 for details.

\hspace{5mm} \textbf{J1502$+$3753} A 5 GHz MERLIN image
shows that the very blue optical/NIR counterpart sits in the center of a very extended, double-lobed source
\citep{bol06}. It is an unlikely candidate for radio absorption studies because the radio source structure is so
extended, with only a weak compact, core component. 

\section{Discussion and Conclusions}

With the goal of finding more high-z molecular absorbers, we have selected 70 strong radio sources with non-elliptical morphology in the optical and obtained NIR images at the APO 3.5m telescope to further characterize their basic nature. Selection from the FIRST survey at $f_{20{\textrm {\scriptsize cm}}}>300$ mJy assures that we are targeting radio-loud AGN. An SDSS associated galaxy classification as non-elliptical gives us the greatest probability of finding plentiful gas, regardless of the reason for the late-type morphologies; e.g., the observed galaxy could either be a very young radio galaxy like a CSO or a foreground galaxy like in a gravitational lens system. In this first paper on ``Invisible AGN'' we have presented a selection method for finding highly-obscured AGN and the optical-NIR data necessary both to confirm their morphology and judge the amount of obscuration present.

We have obtained NIR images of the sample to supply us with morphological information to check the SDSS classification, to extend the SDSS optical SEDs into the near-IR and to search for heavily obscured nuclei. Our NIR observations are consistent morphologically and photometrically with the expectation that many of these sources are heavily obscured systems.  Analysis of the NIR morphology and of the optical/NIR SEDs provides a consistent picture of a heavily obscured, radio-loud AGN in most cases (see detailed discussion in Section 4.2). Also we have made a comparison between the photomeric redshifts and redshifts estimated from the $K$-band magnitudes.  Because radio-loud AGN exhibit a significant ``Hubble diagram'' type correlation out to $z\sim$3, we can use the $K$-band magnitudes as a crude redshift indicator. For nine sources these K-band derived redshifts are significantly greater than their photo-$z$s (see Table 4 and Figure 7), indicating either that the host galaxy is not a giant elliptical or that the SDSS detected galaxy is forground to the AGN. In either case, these are excellent candidates for finding foreground absorption. However, this analysis also makes it clear that it is not advisable to use photometric redshifts of {\it Q} or {\it Q$+$abs}-type sources to conduct an efficient \ion{H}{1} 21~cm absorption-line search. Additionally, Figure 7b shows that even when just the {\it G}-type sources are used, there is a systematic offset ($\Delta z \sim 0.05-0.1$) between the SDSS-derived photometric redshifts and the spectroscopic redshifts on top of the standard statistical errors present in the SDSS photo-$z$ measurement ($\Delta z \sim 0.1$--0.12). While it may be possible to conduct an \ion{H}{1} 21~cm search of sources with {\it G}-type SEDs using photometric redshifts derived using both SDSS+NIR photometry (called $z_{\textrm{{\scriptsize fit}}}$ herein), the statistical errors present (see Figure 7b) are too large to make such a search successful in the presence of the typical RFI environment found at ground-based radio telescopes. Again a more accurate spectroscopic redshift is required.

While the optical/NIR properties described herein are compelling that most of the sources within sample are obscured, 'invisible AGN', it is also important to continue to scrutinize the basic properties of the radio sources against which absorption might be observed. Regardless of the amount of obscuration between us and the nucleus of an AGN, its radio spectrum is unlikely to show strong absorption lines if the radio source contains mostly geometrically-extended flux. This is because we are viewing the large majority of the source flux along unobscured lines-of-sight that avoid going through the host galaxy (and its obscured nucleus) entirely. VLA A-configuration and VLBA maps are in-hand and will provide identification of the most compact radio sources in this sample. These high resolution maps also will be used to make an accurate comparison between the radio source structure and the optical/NIR galaxy positions to identify potential intervening absorption candidates. These maps and their analysis will be presented in Paper 2.  Not only will VLA/VLBA observations discover new examples of CSOs and CSSs but they will also identify those obscured nuclei which are likely to show the strongest absorption due to possessing very compact radio continuum structures.  Hopefully, some of the absorptions we discover will be molecular lines.

\acknowledgments

This project was started in 2006 and 2007 by Fred Hearty during his PhD study. JTS and TY
acknowledge financial support for this work from the National Science Foundation, grant
number: AST-0707480. TY thanks Kyle Willett, Corey Wood, and Brian Keeney for observing assistance. 

{\it Facilities:} \facility{ARC}, \facility{Sloan}, \facility{VLA}, \facility{GBT}.

\clearpage
\begin{deluxetable}{lcccccl}
\tablecaption{Basic Source Data.}
\tablehead{\colhead{Object\tablenotemark{a}} & \colhead{Radio Name} & \colhead{$F_{20\textrm{{\scriptsize cm}}}$} & \colhead{SDSS Name} & \colhead{SDSS $r$} & \colhead{SDSS $z_{\textrm{{\scriptsize phot}}}$} & \colhead{$z_{\textrm{{\scriptsize spec}}}$\tablenotemark{b}} \\ 
\colhead{} & \colhead{} & \colhead{(Jy)} & \colhead{} & \colhead{} & \colhead{} & \colhead{} } 

\startdata
J0000$-$1054$^{\dagger}$	&	PKS 2358$-$111	&	0.40	&	J000057.65$-$105432.1	&	22.01	$\pm$	0.18	&	0.37	$\pm$	0.14	&	  	\\
J0003$-$1053	&	PKS 0001$-$111	&	0.40	&	J000356.30$-$105302.4	&	21.70	$\pm$	0.15	&	0.57	$\pm$	0.12	&	1.474$^*$	\\
J0134$+$0003	&	4C $-$00.11	&	0.89	&	J013412.70$+$000345.2	&	22.74	$\pm$	0.16	&	0.77	$\pm$	0.12	&	0.879	\\
J0249$-$0759	&	PKS 0247$-$08	&	0.62	&	J024935.38$-$075921.2	&	21.63	$\pm$	0.15	&	0.41	$\pm$	0.16	&	  	\\
J0736$+$2954	&	TXS 0733$+$300	&	0.49	&	J073613.67$+$295422.2	&	22.49	$\pm$	0.17	&	0.09	$\pm$	0.05	&	  	\\
J0747$+$4618	&	4C $+$46.16	&	0.52	&	J074743.57$+$461857.7	&	21.38	$\pm$	0.07	&	0.19	$\pm$	0.06	&	2.926	\\
J0749$+$2129	&	TXS 0746$+$216	&	0.42	&	J074948.75$+$212932.9	&	20.94	$\pm$	0.08	&	0.46	$\pm$	0.01	&	  	\\
J0751$+$2716	&	B2 0748$+$27	&	0.60	&	J075141.51$+$271631.7	&	22.23	$\pm$	0.13	&	0.72	$\pm$	0.22	&	0.349\tablenotemark{c}	\\
J0759$+$5312	&	TXS 0755$+$533	&	0.33	&	J075906.52$+$531247.8	&	21.75	$\pm$	0.09	&	0.64	$\pm$	0.24	&	  	\\
J0805$+$1614	&	PKS 0802$+$16	&	0.63	&	J080502.18$+$161404.9	&	20.54	$\pm$	0.06	&	0.54	$\pm$	0.12	&	0.632:$^*$	\\
J0807$+$5327	&	TXS 0803$+$536	&	0.33	&	J080740.73$+$532739.8	&	20.48	$\pm$	0.04	&	0.30	$\pm$	0.12	&	  	\\
J0824$+$5413	&	TXS 0820$+$543	&	0.39	&	J082425.46$+$541349.0	&	19.95	$\pm$	0.05	&	0.46	$\pm$	0.12	&	0.639$^*$	\\
J0834$+$1700$^{\dagger}$	&	4C $+$17.45	&	1.64	&	J083448.22$+$170042.4	&	22.01	$\pm$	0.14	&	0.27	$\pm$	0.15	&	  	\\
J0839$+$2403	&	4C $+$24.18	&	0.66	&	J083957.90$+$240311.4	&	21.15	$\pm$	0.09	&	0.79	$\pm$	0.17	&	  	\\
J0843$+$4215	&	B3 0840$+$424A	&	1.46	&	J084331.63$+$421529.4	&	21.39	$\pm$	0.14	&	0.42	$\pm$	0.15	&	  	\\
J0901$+$0304	&	PKS 0859$+$032	&	0.38	&	J090150.96$+$030423.0	&	18.19	$\pm$	0.01	&	0.29	$\pm$	0.04	&	0.287$^*$	\\
J0903$+$5012	&	4C $+$50.28	&	0.95	&	J090349.81$+$501235.5	&	22.19	$\pm$	0.22	&	0.40	$\pm$	0.22	&	  	\\
J0905$+$4128	&	B3 0902$+$416	&	0.48	&	J090522.18$+$412839.8	&	21.83	$\pm$	0.09	&	0.65	$\pm$	0.16	&	  	\\
J0907$+$0413	&	4C $+$04.30	&	0.70	&	J090750.74$+$041337.3	&	22.57	$\pm$	0.19	&	0.69	$\pm$	0.11	&	  	\\
J0910$+$2419	&	4C $+$24.19	&	0.83	&	J091022.55$+$241919.5	&	20.92	$\pm$	0.06	&	0.74	$\pm$	0.14	&	  	\\
J0915$+$1018	&	TXS 0912$+$105	&	0.35	&	J091512.95$+$101827.4	&	22.01	$\pm$	0.12	&	0.31	$\pm$	0.03	&	  	\\
J0917$+$4725	&	B3 0914$+$476	&	0.35	&	J091727.16$+$472525.6	&	21.82	$\pm$	0.12	&	0.81	$\pm$	0.22	&	  	\\
J0920$+$1753	&	4C $+$18.29	&	1.08	&	J092011.12$+$175324.7	&	22.43	$\pm$	0.16	&	0.70	$\pm$	0.11	&	  	\\
J0920$+$2714	&	B2 0917$+$27B	&	0.46	&	J092045.05$+$271406.6	&	18.04	$\pm$	0.01	&	0.17	$\pm$	0.01	&	0.206$^*$	\\
J0939$+$0304	&	PKS 0937$+$033	&	0.47	&	J093945.16$+$030426.4	&	20.44	$\pm$	0.05	&	0.47	$\pm$	0.02	&	  	\\
J0945$+$2640	&	B2 0942$+$26	&	0.57	&	J094530.97$+$264052.9	&	21.86	$\pm$	0.12	&	0.09	$\pm$	0.03	&	  	\\
J0951$+$1154	&	TXS 0948$+$121	&	0.37	&	J095133.84$+$115459.6	&	21.81	$\pm$	0.09	&	0.60	$\pm$	0.10	&	  	\\
J1008$+$2401	&	B2 1005$+$24	&	0.43	&	J100832.62$+$240119.6	&	22.35	$\pm$	0.16	&	0.82	$\pm$	0.20	&	  	\\
J1010$+$4159	&	B3 1007$+$422	&	0.42	&	J101024.79$+$415933.3	&	20.69	$\pm$	0.07	&	0.49	$\pm$	0.14	&	  	\\
J1019$+$4408	&	B3 1016$+$443	&	0.34	&	J101948.17$+$440824.5	&	22.27	$\pm$	0.18	&	0.09	$\pm$	0.05	&	  	\\
J1023$+$0424	&	TXS 1021$+$046	&	0.33	&	J102337.55$+$042413.9	&	21.62	$\pm$	0.08	&	0.67	$\pm$	0.11	&	  	\\
J1033$+$3935$^{\dagger}$	&	B2 1030$+$39	&	0.41	&	J103322.03$+$393551.0	&	21.45	$\pm$	0.09	&	0.20	$\pm$	0.02	&	1.095	\\
J1034$+$1112	&	TXS 1031$+$114	&	1.21	&	J103405.08$+$111231.6	&	22.21	$\pm$	0.17	&	0.39	$\pm$	0.14	&	  	\\
J1043$+$0537	&	4C $+$05.45	&	0.67	&	J104340.30$+$053712.6	&	22.37	$\pm$	0.17	&	0.53	$\pm$	0.16	&	  	\\
J1045$+$0455	&	TXS 1043$+$051	&	0.38	&	J104551.74$+$045551.1	&	21.62	$\pm$	0.08	&	0.43	$\pm$	0.13	&	  	\\
J1048$+$3457	&	4C $+$35.23	&	1.05	&	J104834.24$+$345724.9	&	20.83	$\pm$	0.04	&	0.39	$\pm$	0.09	&	1.594	\\
J1120$+$2327$^{\dagger}$	&	4C $+$23.27	&	1.38	&	J112043.02$+$232755.2	&	21.81	$\pm$	0.16	&	0.58	$\pm$	0.24	&	1.819	\\
J1125$+$1953	&	PKS 1123$+$201	&	0.43	&	J112555.22$+$195344.0	&	21.18	$\pm$	0.10	&	0.26	$\pm$	0.12	&	  	\\
J1127$+$5743	&	TXS 1124$+$579	&	0.65	&	J112743.71$+$574316.2	&	21.29	$\pm$	0.08	&	0.53	$\pm$	0.02	&	  	\\
J1129$+$5638	&	TXS 1126$+$569	&	0.50	&	J112904.15$+$563844.4	&	21.89	$\pm$	0.20	&	0.67	$\pm$	0.10	&	  	\\
J1142$+$0235	&	TXS 1139$+$028	&	0.38	&	J114206.38$+$023533.5	&	20.90	$\pm$	0.05	&	0.53	$\pm$	0.05	&	  	\\
J1147$+$4818	&	4C $+$48.33	&	0.50	&	J114752.27$+$481849.5	&	22.21	$\pm$	0.20	&	0.17	$\pm$	0.05	&	  	\\
J1148$+$1404	&	TXS 1145$+$143	&	0.33	&	J114825.43$+$140449.3	&	21.66	$\pm$	0.11	&	0.40	$\pm$	0.14	&	  	\\
J1202$+$1207	&	TXS 1200$+$124	&	0.37	&	J120252.09$+$120720.6	&	20.62	$\pm$	0.04	&	0.33	$\pm$	0.13	&	  	\\
J1203$+$4632$^{\dagger}$	&	B3 1200$+$468	&	0.42	&	J120331.80$+$463255.6	&	22.23	$\pm$	0.13	&	0.50	$\pm$	0.12	&	  	\\
J1207$+$5407	&	4C $+$54.26	&	0.60	&	J120714.13$+$540754.5	&	22.11	$\pm$	0.18	&	0.77	$\pm$	0.14	&	  	\\
J1215$+$1730	&	4C $+$17.54	&	1.03	&	J121514.70$+$173002.2	&	21.77	$\pm$	0.10	&	0.07	$\pm$	0.03	&	  	\\
J1228$+$5348	&	4C $+$54.28	&	0.53	&	J122850.60$+$534801.6	&	22.52	$\pm$	0.20	&	0.35	$\pm$	0.19	&	  	\\
J1238$+$0845	&	TXS 1235$+$090	&	0.41	&	J123819.26$+$084500.4	&	21.52	$\pm$	0.10	&	0.75	$\pm$	0.10	&	  	\\
J1300$+$5029	&	TXS 1258$+$507	&	0.38	&	J130041.24$+$502937.0	&	21.55	$\pm$	0.07	&	0.27	$\pm$	0.14	&	1.561	\\
J1312$+$1710	&	TXS 1310$+$174	&	0.34	&	J131235.21$+$171055.7	&	22.52	$\pm$	0.16	&	0.47	$\pm$	0.16	&	  	\\
J1315$+$0222	&	TXS 1312$+$026	&	0.52	&	J131516.93$+$022221.1	&	20.99	$\pm$	0.13	&	0.51	$\pm$	0.17	&	  	\\
J1341$+$1032	&	4C $+$10.36	&	0.69	&	J134104.36$+$103207.3	&	21.99	$\pm$	0.14	&	0.38	$\pm$	0.11	&	  	\\
J1345$+$5846	&	4C $+$59.20	&	0.42	&	J134538.32$+$584654.6	&	22.01	$\pm$	0.17	&	0.58	$\pm$	0.22	&	  	\\
J1347$+$1217$^{\dagger}$	&	4C $+$12.50	&	4.86	&	J134733.35$+$121724.2	&	15.35	$\pm$	0.00	&	0.18	$\pm$	0.05	&	0.122	\\
J1348$+$2415$^{\dagger}$	&	4C $+$24.28	&	0.56	&	J134814.88$+$241550.3	&	22.31	$\pm$	0.18	&	0.21	$\pm$	0.17	&	2.879	\\
J1352$+$3126$^{\dagger}$	&	4C $+$31.43	&	3.70	&	J135217.87$+$312646.4	&	14.12	$\pm$	0.00	&	0.09	$\pm$	0.03	&	0.045	\\
J1354$+$5650	&	4C $+$57.23	&	0.72	&	J135400.10$+$565005.1	&	22.60	$\pm$	0.20	&	0.21	$\pm$	0.08	&	  	\\
J1357$+$0046	&	PKS 1355$+$01	&	2.00	&	J135753.71$+$004633.3	&	22.05	$\pm$	0.11	&	0.57	$\pm$	0.18	&		\\
J1410$+$4850	&	TXS 1408$+$490	&	0.33	&	J141036.04$+$485040.4	&	20.99	$\pm$	0.05	&	0.52	$\pm$	0.09	&	  	\\
J1413$-$0312$^{\dagger}$	&	TXS 1410$-$029	&	0.33	&	J141314.87$-$031227.3	&	12.01	$\pm$	0.00	&	0.12	$\pm$	0.06	&	0.006	\\
J1413$+$1509	&	TXS 1411$+$154	&	0.50	&	J141341.70$+$150939.9	&	20.96	$\pm$	0.09	&	0.43	$\pm$	0.11	&	  	\\
J1414$+$4554	&	B3 1412$+$461	&	0.41	&	J141414.84$+$455448.7	&	20.16	$\pm$	0.03	&	0.48	$\pm$	0.03	&	0.186:	\\
J1415$+$1320$^{\dagger}$	&	PKS 1413$+$135	&	1.18	&	J141558.81$+$132023.7	&	19.05	$\pm$	0.02	&	0.29	$\pm$	0.05	&	0.247	\\
J1421$-$0246$^{\dagger}$	&	4C $-$02.60	&	0.55	&	J142113.54$-$024646.0	&	20.34	$\pm$	0.10	&	0.50	$\pm$	0.03	&	  	\\
J1424$+$1852	&	4C $+$19.47	&	0.70	&	J142409.71$+$185253.5	&	21.49	$\pm$	0.08	&	0.52	$\pm$	0.11	&	  	\\
J1502$+$3753	&	B2 1500$+$38	&	0.34	&	J150234.77$+$375353.0	&	20.50	$\pm$	0.06	&	0.40	$\pm$	0.09	&	  	\\
J1504$+$5438	&	TXS 1503$+$548	&	0.38	&	J150451.19$+$543839.7	&	20.29	$\pm$	0.05	&	0.53	$\pm$	0.10	&	0.621	\\
J1504$+$6000	&	TXS 1502$+$602	&	1.55	&	J150409.21$+$600055.7	&	20.00	$\pm$	0.02	&	0.47	$\pm$	0.16	&	1.024$^*$	\\
J1523$+$1332	&	4C $+$13.54	&	0.35	&	J152321.74$+$133229.2	&	21.76	$\pm$	0.12	&	0.46	$\pm$	0.18	&	  	\\
J1527$+$3312	&	B2 1525$+$33	&	0.32	&	J152750.89$+$331253.0	&	22.04	$\pm$	0.10	&	0.44	$\pm$	0.14	&	  	\\
J1528$-$0213	&	PKS 1525$-$020	&	0.47	&	J152821.99$-$021319.0	&	21.75	$\pm$	0.15	&	0.63	$\pm$	0.07	&	  	\\
J1548$+$0808	&	TXS 1545$+$082	&	0.63	&	J154809.05$+$080834.7	&	21.78	$\pm$	0.14	&	0.65	$\pm$	0.12	&	  	\\
J1551$+$6405	&	TXS 1550$+$642	&	0.68	&	J155128.19$+$640537.8	&	22.10	$\pm$	0.15	&	0.47	$\pm$	0.15	&	  	\\
J1559$+$4349	&	4C $+$43.36	&	0.75	&	J155931.21$+$434916.6	&	21.11	$\pm$	0.07	&	0.42	$\pm$	0.09	&	1.232$^*$	\\
J1604$+$6050	&	TXS 1603$+$609	&	0.59	&	J160427.40$+$605055.5	&	21.44	$\pm$	0.10	&	0.59	$\pm$	0.05	&	  	\\
J1616$+$2647	&	PKS 1614$+$26	&	1.41	&	J161638.32$+$264701.3	&	22.46	$\pm$	0.17	&	0.44	$\pm$	0.03	&	  	\\
J1625$+$4134	&	4C $+$41.32	&	1.72	&	J162557.68$+$413440.8	&	22.35	$\pm$	0.16	&	0.58	$\pm$	0.18	&	2.550	\\
J1629$+$1342$^{\dagger}$	&	4C $+$13.60	&	0.71	&	J162948.42$+$134205.7	&	22.34	$\pm$	0.18	&	0.44	$\pm$	0.18	&	  	\\
J1633$+$4700	&	4C $+$47.43	&	0.46	&	J163315.02$+$470016.9	&	21.44	$\pm$	0.12	&	0.57	$\pm$	0.19	&	  	\\
J1724$+$3852	&	B2 1722$+$38	&	0.37	&	J172400.53$+$385226.7	&	22.14	$\pm$	0.12	&	0.11	$\pm$	0.05	&	  	\\
J2203$-$0021	&	4C $-$00.79	&	0.61	&	J220358.31$-$002148.1	&	21.21	$\pm$	0.13	&	0.57	$\pm$	0.10	&	0.729$^*$	\\

\enddata

\tablenotetext{a}{Objects noted with superscript ``$\dagger$'' are removed from final sample because the source is : (1) re-evaluated morphologically to be early-type using SDSS-DR8 (J0000$-$1054, J1033$+$3935, J1203$+$4632, J1421$-$0246, J1629$+$1342); (2) nearby at $z < 0.1$ (J1352$+$3126, J1413$-$0312); (3) a well-studied source (molecular absorption system J1415$+$1320, alignment-effect systems J1120$+$2327 \& J1348$+$2415, low-$z$ CSO J1347$+$1217); or (5) aligned with one lobe of a double-lobe radio galaxy (J0834$+$1700).}
\tablenotetext{b}{Redshifts noted with superscript ``*'' were obtained recently by us using DIS II or TripleSpec at the APO 3.5m telescope (see Figures \ref{fg:DIS} \& \ref{fg:TSPEC}). Others come from the NASA/NED database. Redshifts noted with ``:'' are uncertain.}
\tablenotetext{c}{Gravitational lensing system with background quasar at $z=3.2$ and foreground galaxy at $z=0.349$.}

\end{deluxetable}

\clearpage
\begin{deluxetable}{crrrcccc}
\tablecaption{NIR photometry and optical-NIR SED.}
\tablehead{\colhead{Object} & \colhead{$J$} & \colhead{$H$} & \colhead{$K_{\textrm{{\scriptsize short}}}$} & \colhead{Observation} & \colhead{SED} & \colhead{$z_{\textrm{{\scriptsize fit}}}$} & \colhead{Hubble}\\
\colhead{} & \colhead{} & \colhead{} & \colhead{} & \colhead{Date} & \colhead{Type\tablenotemark{a}} & \colhead{} & \colhead{Type\tablenotemark{b}}
} 

\startdata
	J0003$-$1053	&		17.69	$\pm$	0.28	&		16.46	$\pm$	0.28	&		16.19	$\pm$	0.31	&	UT071213	&	Q+abs	&		&		\\
	J0134$+$0003	&		18.88	$\pm$	0.38	&		17.76	$\pm$	0.39	&		16.91	$\pm$	0.39	&	UT071213	&	G	&	0.92	&	Sa	\\
	J0249$-$0759	&		19.19	$\pm$	0.21	&		18.30	$\pm$	0.22	&		17.40	$\pm$	0.21	&	UT081102	&	G+Q	&		&		\\
	J0736$+$2954	&		19.62	$\pm$	0.27	&		18.57	$\pm$	0.27	&		17.60	$\pm$	0.27	&	UT081228	&	G+Q	&		&		\\
	J0747$+$4618	&		19.40	$\pm$	0.14	&					&		17.95	$\pm$	0.14	&	UT091228	&	Q+abs	&		&		\\
	J0749$+$2129	&		18.22	$\pm$	0.13	&					&		16.37	$\pm$	0.13	&	UT080120	&	G	&	0.52	&	S0	\\
	J0751$+$2716	&		19.84	$\pm$	0.18	&		19.41	$\pm$	0.18	&		18.18	$\pm$	0.17	&	UT070304	&	G	&	0.42	&	Sc	\\
	J0759$+$5312	&		19.66	$\pm$	0.17	&		18.65	$\pm$	0.17	&		18.09	$\pm$	0.17	&	UT080218	&	G+Q	&		&		\\
	J0805$+$1614	&		18.12	$\pm$	0.11	&		16.78	$\pm$	0.11	&		15.87	$\pm$	0.12	&	UT080218	&	G	&	0.67	&	Sc	\\
	J0807$+$5327	&		19.13	$\pm$	0.09	&					&		17.73	$\pm$	0.09	&	UT080120	&	Q+abs	&		&		\\
	J0824$+$5413	&		17.67	$\pm$	0.09	&					&		15.81	$\pm$	0.10	&	UT080120	&	G+Q	&		&		\\
	J0839$+$2403	&		18.83	$\pm$	0.15	&		18.60	$\pm$	0.15	&		17.63	$\pm$	0.15	&	UT070304	&	G+Q	&		&		\\
	J0843$+$4215	&		18.59	$\pm$	0.24	&					&		16.51	$\pm$	0.25	&	UT080120	&	G	&	0.60	&	Sb	\\
	J0901$+$0304	&		16.17	$\pm$	0.03	&		15.18	$\pm$	0.02	&		14.45	$\pm$	0.03	&	UT080120	&	G	&	0.21	&	Sc	\\
	J0903$+$5012	&		19.16	$\pm$	0.35	&		17.95	$\pm$	0.35	&		17.44	$\pm$	0.35	&	UT080218	&	G+Q	&		&		\\
	J0905$+$4128	&		19.46	$\pm$	0.17	&		19.14	$\pm$	0.17	&		17.48	$\pm$	0.17	&	UT080218	&	G+Q	&		&		\\
	J0907$+$0413	&		19.32	$\pm$	0.33	&		18.59	$\pm$	0.33	&		17.51	$\pm$	0.33	&	UT070304	&	G	&	1.10	&	Sc	\\
	J0910$+$2419	&		18.04	$\pm$	0.10	&					&		17.19	$\pm$	0.11	&	UT080120	&	G+Q	&		&		\\
	J0915$+$1018	&		20.02	$\pm$	0.23	&		19.31	$\pm$	0.23	&		18.33	$\pm$	0.23	&	UT081228	&	G+Q	&		&		\\
	J0917$+$4725	&					&		18.54	$\pm$	0.22	&		17.61	$\pm$	0.22	&	UT081228	&	G+Q	&		&		\\
	J0920$+$1753	&		19.00	$\pm$	0.31	&		18.49	$\pm$	0.31	&		17.53	$\pm$	0.31	&	UT090116	&	G	&	0.68	&	Sc	\\
	J0920$+$2714	&		15.91	$\pm$	0.04	&		14.94	$\pm$	0.05	&		14.36	$\pm$	0.06	&	UT060611	&	G	&	0.23	&	S0	\\
	J0939$+$0304	&		17.80	$\pm$	0.09	&		17.01	$\pm$	0.09	&		16.46	$\pm$	0.10	&	UT060611	&	G	&	0.34	&	S0	\\
	J0945$+$2640	&		19.33	$\pm$	0.21	&					&		18.63	$\pm$	0.21	&	UT080418	&	Q	&		&		\\
	J0951$+$1154	&		18.86	$\pm$	0.17	&		18.24	$\pm$	0.17	&		17.37	$\pm$	0.17	&	UT081228	&	G	&	0.65	&	Sc	\\
	J1008$+$2401	&		19.41	$\pm$	0.35	&		18.76	$\pm$	0.36	&		17.80	$\pm$	0.36	&	UT081228	&	G	&	0.62	&	Sc	\\
	J1010$+$4159	&		17.79	$\pm$	0.12	&		17.09	$\pm$	0.12	&		16.79	$\pm$	0.13	&	UT060611	&	G	&	0.71	&	Sc	\\
	J1019$+$4408	&		20.05	$\pm$	0.27	&		19.07	$\pm$	0.27	&		17.95	$\pm$	0.27	&	UT081228	&	Q+abs	&		&		\\
	J1023$+$0424	&		18.74	$\pm$	0.15	&		18.56	$\pm$	0.15	&		17.63	$\pm$	0.15	&	UT060611	&	G	&	0.71	&	Sc	\\
	J1034$+$1112	&		20.19	$\pm$	0.29	&					&		17.94	$\pm$	0.29	&	UT090116	&	G+Q	&		&		\\
	J1043$+$0537	&		19.75	$\pm$	0.29	&					&		17.96	$\pm$	0.29	&	UT090116	&	G	&	0.42	&	E	\\
	J1045$+$0455	&		19.67	$\pm$	0.15	&		18.76	$\pm$	0.15	&		18.53	$\pm$	0.15	&	UT081228	&	Q+abs	&		&		\\
	J1048$+$3457	&		18.39	$\pm$	0.08	&		17.74	$\pm$	0.13	&		17.31	$\pm$	0.12	&	UT060611	&	Q+abs	&		&		\\
	J1125$+$1953	&		20.13	$\pm$	0.17	&					&	$>$	19.08	$\pm$	0.26	&	UT090116	&	Q	&		&		\\
	J1127$+$5743	&		18.78	$\pm$	0.12	&		17.83	$\pm$	0.12	&		17.12	$\pm$	0.15	&	UT060611	&	G	&	0.49	&	Sb	\\
	J1129$+$5638	&		18.57	$\pm$	0.31	&		18.02	$\pm$	0.31	&		17.38	$\pm$	0.31	&	UT081228	&	G	&	0.64	&	Sa	\\
	J1142$+$0235	&		18.90	$\pm$	0.09	&		17.30	$\pm$	0.08	&		17.04	$\pm$	0.09	&	UT060611	&	G	&	0.37	&	Sa	\\
	J1147$+$4818	&		19.92	$\pm$	0.28	&					&		17.23	$\pm$	0.28	&	UT090116	&	Q	&		&		\\
	J1148$+$1404	&		19.35	$\pm$	0.20	&					&		17.72	$\pm$	0.20	&	UT090116	&	G	&	0.54	&	Sc	\\
	J1202$+$1207	&		19.36	$\pm$	0.08	&					&		17.27	$\pm$	0.10	&	UT090116	&	G+Q	&		&		\\
	J1207$+$5407	&		18.92	$\pm$	0.31	&		18.26	$\pm$	0.31	&		17.58	$\pm$	0.31	&	UT081228	&	G	&	0.61	&	Sc	\\
	J1215$+$1730	&		19.78	$\pm$	0.16	&		19.04	$\pm$	0.16	&		18.11	$\pm$	0.17	&	UT080120	&	Q+abs	&		&		\\
	J1228$+$5348	&		19.31	$\pm$	0.36	&					&		19.03	$\pm$	0.36	&	UT090116	&	G+Q	&		&		\\
	J1238$+$0845	&		17.93	$\pm$	0.16	&		17.58	$\pm$	0.16	&		16.88	$\pm$	0.17	&	UT060611	&	G	&	0.65	&	Sc	\\
	J1300$+$5029	&		19.27	$\pm$	0.13	&					&		17.60	$\pm$	0.13	&	UT080517	&	Q	&		&		\\
	J1312$+$1710	&		19.55	$\pm$	0.37	&		19.02	$\pm$	0.37	&		18.18	$\pm$	0.37	&	UT080218	&	G	&	0.63	&	Sc	\\
	J1315$+$0222	&					&		16.91	$\pm$	0.23	&		16.30	$\pm$	0.23	&	UT060611	&	G	&	0.85	&	Sc	\\
	J1341$+$1032	&		20.02	$\pm$	0.25	&	$>$	19.24	$\pm$	0.39	&		17.79	$\pm$	0.25	&	UT081228	&	Q+abs	&		&		\\
	J1345$+$5846	&		18.86	$\pm$	0.30	&					&		18.25	$\pm$	0.30	&	UT091228	&	G+Q	&		&		\\
	J1354$+$5650	&		19.82	$\pm$	0.30	&					&		18.14	$\pm$	0.30	&	UT080517	&	G	&	0.65	&	Sc	\\
	J1357$+$0046	&		19.77	$\pm$	0.19	&		18.63	$\pm$	0.19	&		18.09	$\pm$	0.19	&	UT081228	&	G+Q	&		&		\\
	J1410$+$4850	&					&		18.39	$\pm$	0.10	&		17.08	$\pm$	0.10	&	UT060630	&	G+Q	&		&		\\
	J1413$+$1509	&		18.63	$\pm$	0.16	&					&		17.37	$\pm$	0.18	&	UT080120	&	G	&	0.22	&	S0	\\
	J1414$+$4554	&		17.62	$\pm$	0.06	&		16.63	$\pm$	0.06	&		16.15	$\pm$	0.06	&	UT070603	&	G	&	0.38	&	S0	\\
	J1424$+$1852	&		19.26	$\pm$	0.17	&		18.97	$\pm$	0.18	&		17.77	$\pm$	0.19	&	UT080218	&	Q+abs	&		&		\\
	J1502$+$3753	&		18.37	$\pm$	0.10	&		17.87	$\pm$	0.09	&					&	UT060630	&	Q+abs	&		&		\\
	J1504$+$5438	&		18.09	$\pm$	0.09	&		17.00	$\pm$	0.09	&		16.15	$\pm$	0.09	&	UT080218	&	G	&	0.51	&	Sc	\\
	J1504$+$6000	&		18.06	$\pm$	0.07	&		18.46	$\pm$	0.05	&		17.60	$\pm$	0.06	&	UT070528	&	Q	&		&		\\
	J1523$+$1332	&					&		18.51	$\pm$	0.21	&		18.26	$\pm$	0.21	&	UT070528	&	Q+abs	&		&		\\
	J1527$+$3312	&					&					&		18.11	$\pm$	0.21	&	UT080517	&	G	&	0.32	&	Sb	\\
	J1528$-$0213	&					&		17.62	$\pm$	0.25	&		16.75	$\pm$	0.25	&	UT070528	&	G	&	0.54	&	E	\\
	J1548$+$0808	&		18.77	$\pm$	0.25	&		17.72	$\pm$	0.25	&		16.83	$\pm$	0.25	&	UT070528	&	G	&	0.51	&	E	\\
	J1551$+$6405	&					&					&		18.88	$\pm$	0.27	&	UT110518	&	G+Q	&		&		\\
	J1559$+$4349	&		18.55	$\pm$	0.13	&		17.70	$\pm$	0.12	&		16.78	$\pm$	0.12	&	UT060630	&	Q+abs	&		&		\\
	J1604$+$6050	&		18.94	$\pm$	0.18	&		17.74	$\pm$	0.18	&		16.89	$\pm$	0.18	&	UT080218	&	G	&	0.42	&	S0	\\
	J1616$+$2647	&					&		18.59	$\pm$	0.27	&		18.10	$\pm$	0.27	&	UT070528	&	G	&	0.72	&	Sc	\\
	J1625$+$4134	&		19.21	$\pm$	0.26	&		17.98	$\pm$	0.26	&		16.80	$\pm$	0.26	&	UT060630	&	Q	&		&		\\
	J1633$+$4700	&		19.10	$\pm$	0.20	&		18.49	$\pm$	0.20	&		17.73	$\pm$	0.20	&	UT080218	&	Q+abs	&		&		\\
	J1724$+$3852	&		19.96	$\pm$	0.09	&		18.50	$\pm$	0.09	&		18.72	$\pm$	0.09	&	UT081102	&	Q+abs	&		&		\\
	J2203$-$0021	&		18.54	$\pm$	0.18	&		17.36	$\pm$	0.18	&		16.85	$\pm$	0.18	&	UT081102	&	G	&	0.71	&	Sc	\\

\enddata

\tablenotetext{a}{The optical-NIR SED is classified as $G$ if it drops at the blue end, $Q$ if it keeps increasing or $Q+abs$ if it flattens out at the blue end, and $G+Q$ if there is flux excess at the red end on top of $Q$ feature at the blue end. See Section \ref{sec:SED} for details.}
\tablenotetext{b}{Sources classified as $G$ are fit with template spectra (Figure \ref{fg:template}) to estimate the Hubble type of the host galaxy. See section 4.2 for details.}

\end{deluxetable}

\clearpage
\begin{deluxetable}{ccccl}
\rotate
\tablecaption{New Spectroscopic Redshifts from the Apache Point 3.5m Telescope.}
\tablehead{\colhead{Object} & \colhead{$z_{\textrm{{\scriptsize spec}}}$} & \colhead{$z_{\textrm{{\scriptsize phot}}}$} & \colhead{Instrument Used} & \colhead{Lines Seen} 
} 

\startdata
J0003$-$1053	&	 1.474$\pm$0.001 	&	 0.42 	&	 TripleSpec 	&	[\ion{O}{3}], H$\alpha$, [\ion{N}{2}]\\
J0805$+$1614	&	 0.632:$\pm$0.002 	&	 0.49 	&	 TripleSpec 	&	H$\alpha$, \ion{He}{1} 1.08$\mu$; lines are weak \& need confirmation\\
J0824$+$5413	&	 0.6385$\pm$0.0003 	&	 0.46 	&	 DIS~II 	&	[\ion{O}{2}], \ion{Ca}{2} H\&K, H$\beta$, [\ion{O}{3}] \\
	&	            	&	      	&	 TripleSpec 	&	H$\alpha$, [\ion{N}{2}], [\ion{S}{2}], [\ion{S}{3}]: 0.95$\mu$ \\
J0901$+$0304	&	 0.2872$\pm$0.0001 	&	 0.29 	&	 DIS~II 	&	[\ion{O}{2}], H$\beta$, [\ion{O}{3}], [\ion{O}{1}], H$\alpha$, [\ion{N}{2}], [\ion{S}{2}] \\
	&	            	&	      	&	        	&	also detected in \ion{H}{1} @$z$=(0.2869, 0.2886); see Paper 3\\
J0920$+$2714	&	 0.2064$\pm$0.0002 	&	 0.17 	&	 DIS~II 	&	\ion{Ca}{2} H\&K, G-band, H$\alpha$, [\ion{N}{2}], [\ion{S}{2}] \\
	&	            	&	      	&	        	&	also detected in \ion{H}{1} @$z$=0.2067 (Paper 3) \\
J1504$+$6000	&	 1.024$\pm$0.001 	&	 0.67 	&	 DIS~II 	&	\ion{C}{3}], \ion{C}{2}]2326\AA, [\ion{Ne}{5}], [\ion{O}{2}], [\ion{Ne}{3}] 3869, 3967\AA;  \\
	&	            	&	      	&	        	&	slightly revises Burbidge (1970) redshift of $z$=1.022\\
J1559$+$4349	&	 1.232$\pm$0.001 	&	 0.42 	&	 TripleSpec 	&	[\ion{O}{3}], H$\alpha$, [\ion{N}{2}], [\ion{S}{3}] 0.95$\mu$, \ion{He}{1} 1.08$\mu$: \\
J2203$-$0021	&	 0.729$\pm$0.001 	&	 0.57 	&	 TripleSpec 	&	H$\alpha$, [\ion{N}{2}], [\ion{S}{3}] 0.95$\mu$ \\

\enddata

\end{deluxetable}

\clearpage
\begin{deluxetable}{cccccc}
\tablecaption{Sources with Discrepant Redshift Estimates.}
\tablehead{\colhead{Object} & \colhead{$z_{\textrm{{\scriptsize fit}}}$\tablenotemark{a}} & \colhead{Hubble} & \colhead{$K_{{\textrm {\scriptsize short}}}$\tablenotemark{c}} & \colhead{$z_{\textrm{{\scriptsize K}}}$} & \colhead{$\Delta z$} \\ 
\colhead{} & \colhead{} & \colhead{Type\tablenotemark{b}} & \colhead{} & \colhead{} & \colhead{}
} 

\startdata
J1008$+$2401	&	0.62	&	Sc	&	17.80	&	1.25	&	0.63	\\
J1043$+$0537	&	0.42	&	E	&	17.96	&	1.35	&	0.93	\\
J1142$+$0235	&	0.37	&	Sa	&	17.04	&	0.85	&	0.48	\\
J1148$+$1404	&	0.54	&	Sc	&	17.72	&	1.20	&	0.66	\\
J1312$+$1710	&	0.63	&	Sc	&	18.18	&	1.52	&	0.89	\\
J1354$+$5650	&	0.65	&	Sc	&	18.14	&	1.49	&	0.84	\\
J1413$+$1509	&	0.22	&	S0	&	17.37	&	1.00	&	0.78	\\
J1527$+$3312	&	0.32	&	Sb	&	18.11	&	1.46	&	1.14	\\
J1616$+$2647	&	0.72	&	Sc	&	18.10	&	1.46	&	0.74	\\

\enddata

\tablenotetext{a,b,c}{Same as in Table 2.}

\end{deluxetable}

\begin{thebibliography}{}
\bibitem[Becker et al.(1995)]{bec95}
Becker, R. H., White, R. L., \& Helfand, D. J. 1995, \apj, 450, 559
\bibitem[Begelman(1996)]{beg96}
Begelman, M. 1996, in Cygnus A: Study of a Radio Galaxy, ed. C. L. Carilli \& D. E. Harris (Cambridge: Cambridge Univ. Press), 209
\bibitem[Bolton et al.(2006)]{bol06}
Bolton, R. C., Chandler, C. J., Cotter, G., Pearson, T. J., Pooley, G. G., Readhead, A. C. S., Riley, J. M., \& Waldram, E. M. 2006, \mnras, 367, 323
\bibitem[Budav\'{a}ri et al.(2000)]{bud00}
Budav\'{a}ri, T., Szalay, A. S., Connolly, A. J., Csabai, I., \& Dickinson, M., 2000 \aj, 120, 1588
\bibitem[Carilli et al.(1992)]{car92}
Carilli, C. L., Perlman, E. S., \& Stocke, J. T. 1992, \apjl, 400, L13
\bibitem[Chand et al.(2006)]{cha06}
Chand, H., Srianand, R., Petitjean, P., Aracil, B., Quast, R., \& Reimers, D. 2006, \aap, 451, 45
\bibitem[Chandola et al.(2011)]{cha11}
Chandola, Y., Sirothia, S. K., \& Saikia, D. J. 2011, \mnras, 418, 1787
\bibitem[Csabai et al.(2007)]{csa07}
Csabai, I., Dobos, L., Trencs\'{e}ni, M., Herczegh, G., J\'{o}zsa, P., Purger, N., Budav\'{a}ri, T., Szalay, A. S. 2007, Astronomische Nachrichten 328, 852
\bibitem[Csabai et al.(2003)]{csa03}
Csabai, I., et al., 2003, \aj, 125, 580
\bibitem[Curran \& Whiting(2010)]{cur10}
Curran, S. J., \& Whiting, M. T. 2010, \apj, 712, 303
\bibitem[Curran et al.(2008)]{cur08}
Curran, S. J., Whiting, M. T., Wiklind, T., Webb, J. K., Murphy, M. T., \& Purcell, C. R. 2008, \mnras, 391, 765
\bibitem[Curran et al.(2006)]{cur06}
Curran, S. J., Whiting, M. T., Murphy, M. T., Webb, J. K., Longmore, S. N., Pihlstr\"{o}m, Y. M., Athreya, R., \& Blake, C. 2006, \mnras, 371, 431
\bibitem[Curran et al.(2011a)]{cur11a}
Curran, S. J., et al. 2011, \mnras, 413, 1165
\bibitem[Curran et al.(2011b)]{cur11b}
Curran, S. J., et al. 2011, \mnras, 416, 2143
\bibitem[Darling(2003)]{dar03}
Darling, J. 2003, \prl, 91, 11301
\bibitem[Darling(2004)]{dar04}
Darling, J. 2004, \apj, 612, 58
\bibitem[de Vries et al.(1998)]{dev98}
de Vries, W. H., O'Dea, C. P., Perlman, E., Baum, S. A., Lehnert, M. D., Stocke, J., Rector, T., \& Elston, R. 1998, \apj, 503, 138
\bibitem[Drinkwater et al.(1997)]{dri97}
Drinkwater, M. J., et al. 1997, \mnras, 284, 85	
\bibitem[Falco et al.(1998)]{fal98}
Falco, E. E., Kochanek, C. S., \& Mu\~{n}oz, J. A. 1998, \apj, 494, 47
\bibitem[Gaensler(2007)]{gae07}
Gaensler, B. 2007, in From Planets to Dark Energy: the Modern Radio Universe, 66
\bibitem[Gerin et al.(1997)]{ger97}
Gerin, M., Phillips, T. G., Benford, D. J., Young, K. H., Menten, K. M., \& Frye, B. 1997, \apjl, 488, L31 
\bibitem[Glikman et al.(2006)]{gli06}
Glikman, E., Helfand, D. J., \& White, R. L. 2006, \apj, 640, 579
\bibitem[Gupta et al.(2009)]{gup09}
Gupta, N., Srianand, R., Petitjean, P., Noterdaeme, P., \& Saikia, D. J. 2009, \mnras, 398, 201
\bibitem[Gupta et al.(2006)]{gup06}
Gupta, N., Salter, C. J., Saikia, D. J., Ghosh, T., \& Jeyakumar, S. 2006, \mnras, 373, 972
\bibitem[Hearty et al.(2005)]{hea05}
Hearty, F., et al. 2005, \procspie, 5904, 199
\bibitem[Kanekar et al.(2005)]{kan05}
Kanekar, N., et al. 2005, \prl, 95, 261301
\bibitem[Kanekar et al. (2012)]{kan12}
Kanekar, N., Langston, G. I., Stocke, J. T., Carilli, C. L., \& Menten, K. M. 2012, \apjl, 746, L16
\bibitem[Kanekar et al.(2004)]{kan04}
Kanekar, N., Chengalur, J. N., \& Ghosh, T. 2004, \prl, 93, 51302
\bibitem[Lehar et al.(1997)]{leh97}
Lehar, J., et al. 1997, \aj, 114, 48
\bibitem[Lupton et al.(2001)]{lup01}
Lupton, R., Gunn, J. E., Ivezi\'{c}, Z., Knapp, G. R., Kent, S., \& Yasuda, N. 2001, {\it Astronomical Data Analysis Software and Systems X} (ASP Conf. Ser. 238), ed. F. R. Harnden Jr., F. A. Primini, \& H. E. Payne (San Francisco, CA: ASP), 269
\bibitem[Mannucci et al.(2001)]{man01}
Mannucci, F., Basile, F., Poggianti, B. M., Cimatti, A., Daddi, E., Pozzetti, L., \& Vanzi, L. 2001, \mnras, 326, 745
\bibitem[McCarthy (1991)]{mcc91}
McCarthy, P. J. 1991, \aj, 102, 518
\bibitem[Mcleod(2006)]{mcl06}
McLeod, K. K. 2006, in Planets to Cosmology: Essential Science in the Final Years of Hubble Space Telescope, ed. M. Livio \& S. Casertano, (Cambridge: Cambridge Univ. Press), 73
\bibitem[Menten & Reid(1996)]{men96}
Menten, K. M., \& Reid, M. J. 1996, \apjl, 465, L99
\bibitem[Mihos \& Hernquist(1996)]{mih96}
Mihos, J. C., \& Hernquist L. 1996, \apj, 464, 641
\bibitem[Narayan \& Schneider(1990)]{nar90}
Narayan, R., \& Schneider, P. 1990, \mnras, 243, 192
\bibitem[Noterdaeme et al.(2010)]{not10}
Noterdaeme, P., Petitjean, P., Ledoux, C., L\'{o}pez, S., Srianand, R., \& Vergani, S. D. 2010, \aap, 523, 80
\bibitem[Oke \& Gunn(1982)]{oke82}
Oke, J. B., \& Gunn, J. E. 1982, \pasp, 94, 586
\bibitem[Perlman et al.(1996)]{per96}
Perlman, E. S., Carilli, C. L., Stocke, J. T., \& Conway, J. 1996, \aj, 111, 1839
\bibitem[Perlman et al.(2001)]{per01}
Perlman, E. S., Stocke, J. T., Conway, J., \& Reynolds, C. 2001, \aj, 122, 536
\bibitem[Srianand et al.(2008)]{sri08}
Srianand, R., Noterdaeme, P., Ledoux, C., \& Petitjean, P. 2008, \aap, 482, L39
\bibitem[Stocke et al.(1992)]{sto92}
Stocke, J. T., Wurtz, R., Wang, Q., Elston, R., \& Jannuzi, B. T. 1992, \apjl, 400, L17
\bibitem[Stoughton et al.(2002)]{sto02}
Stoughton, C., et al. 2002, \aj, 123, 485
\bibitem[Strateva et al.(2001)]{str01}
Strateva, I., et al. 2001, \aj, 122, 1861
\bibitem[Urry \& Padovani(1995)]{urr95}
Urry, C. M., \& Padovani, P. 1995, \pasp, 107, 803
\bibitem[van Breugel et al.(1998)]{van98}
van Breugel, W. J. M., Stanford, S. A., Spinrad, H., Stern, D., \& Graham, J. R. 1998, \apj, 502, 614
\bibitem[van Gorkom et al.(1989)]{van89}
van Gorkom, J. H., Knapp, G. R., Ekers, R. D., Ekers, D. D., Laing, R. A., \& Polk, K. S. 1989, \aj, 97, 708
\bibitem[Vanden Berk et al.(2001)]{van01}
Vanden Berk, D. E., et al. 2001, \aj, 122, 549
\bibitem[Vermeulen et al.(2003)]{ver03}
Vermeulen, R. C., et al. 2003, \aap, 404, 861
\bibitem[Wiklind \& Combes(1994)]{wik94}
Wiklind, T., \& Combes, F. 1994, \aap, 286, L9   
\bibitem[Wiklind \& Combes(1995)]{wik95}
Wiklind, T., \& Combes, F. 1995, \aap, 299, 382   
\bibitem[Wiklind \& Combes(1996a)]{wik96a}
Wiklind, T., \& Combes, F. 1996a, \nat, 379, 139
\bibitem[Wiklind \& Combes(1996b)]{wik96b}
Wiklind, T., \& Combes, F. 1996b, \aap, 328, 48 
\bibitem[Wiklind \& Combes(1997)]{wik97}
Wiklind, T., \& Combes, F. 1997, \aap, 315, 86   
\bibitem[Willett et al.(2010)]{wil10}
Willett, K. W., Stocke, J. T., Darling, J., \& Perlman, E. S. 2010, \apj, 713, 1393
\bibitem[Willott et al.(2003)]{wil03} 
Willott, C. J., Rawlings, S., Jarvis, M. J., \& Blundell, K. M. 2003, \mnras, 339, 173
\bibitem[Wilson \& Colbert(1995)]{wil95}
Wilson, A. S., \& Colbert, E. J. M. 1995, \apj, 438, 62
\bibitem[Wilson et al.(2004)]{wil04}
Wilson, J. C., et al. 2004, \procspie, 5492, 1295
\bibitem[York et al.(2000)]{yor00}
York, D., et al. 2000, \aj, 120, 1579
\bibitem[Zirm et al.(2003)]{zir03}
Zirm, A. W., Dickinson, M., Dey, A. 2003, \apj, 585, 90 
\end{thebibliography}
\end{document}